\documentclass[%
aip,
amsmath,amssymb,
reprint,%
]{revtex4-1}

\usepackage{graphicx}% Include figure files
\usepackage{dcolumn}% Align table columns on decimal point
\usepackage{bm}% bold math
\usepackage[utf8]{inputenc}
\usepackage[T1]{fontenc}
\usepackage{mathptmx}
\usepackage{etoolbox}

\usepackage{scrextend}
\usepackage[colorlinks=true,
						allcolors=blue]{hyperref}
\usepackage{multirow}
\usepackage[version=4]{mhchem}
\usepackage{booktabs}
\usepackage{amsmath,amssymb}
\usepackage{siunitx}
\DeclareSIUnit\monolayer{ML}
\DeclareSIUnit\torr{torr}
\DeclareSIPrefix\micro{\text{\textmu}}{-3}

\newcommand{\nocontentsline}[3]{}

\newcommand\stoptoc{\let\addcontentsline\nocontentsline}

\graphicspath{{figures}}
\makeatletter
\def\@email#1#2{%
	\endgroup
	\patchcmd{\titleblock@produce}
	{\frontmatter@RRAPformat}
	{\frontmatter@RRAPformat{\produce@RRAP{*#1\href{mailto:#2}{#2}}}\frontmatter@RRAPformat}
	{}{}
}%
\makeatother

\makeatletter
\renewcommand\@makefnmark{\hbox{\@textsuperscript{\normalfont\@thefnmark)}}}
\makeatother

\begin{document}
	
	\preprint{AIP/123-QED}
	
	\title[]{Growth of quantum dots by droplet etching epitaxy in molecular beam epitaxy: theory, practice, and review.}
	% Force line breaks with \\
	\author{D. Gossink}
	\author{U. S. Sainadh}
	\author{G. S. Solomon}
	\email{glenn.solomon@adelaide.edu.au}
	\affiliation{Department of Physics, Adelaide University, Adelaide, SA AU 5005}
	\date{\today}% It is always \today, today,
	
	\begin{abstract}
		The following article has been submitted to Applied Physics Reviews. After it is published, it will be found at \url{pubs.aip.org/aip/apr}.\\
		
		\noindent GaAs quantum dots grown by droplet etching epitaxy are high-quality solid-state sources of quantum light. Despite implementation in devices that exploit quantum phenomenon, a comprehensive review on the crystal growth of quantum dots grown by droplet etching epitaxy is absent, unlike for other quantum dot growth techniques such as the related droplet epitaxy method or Stranski-Krastanov growth of InAs quantum dots. This review presents a detailed overview of the droplet etching epitaxy growth technique in the molecular beam epitaxy environment, with emphasis on the growth parameters necessary to realize high-quality quantum dots. We systematically cover the three main phases of droplet etching epitaxy---droplet deposition, droplet etching, and nanohole regrowth---and relate experimental results to theories on crystal growth. The review concludes with an introduction to GaAs quantum dot photoluminescence and the extension of droplet etching epitaxy beyond the AlGaAs/GaAs material system.
	\end{abstract}
	
	\maketitle
	
	{\hypersetup{linkcolor=black}
	\tableofcontents}
	\section{Introduction}
	The rapid growth of the fields of quantum communication, quantum computing, and quantum metrology in the last two decades has set the stage for the ``second quantum revolution".~\cite{dowlingQuantumTechnologySecond2003, deutschHarnessingPowerSecond2020, aspectSecondQuantumRevolution2023} As opposed to the ``first quantum revolution'', in which the quantized properties of matter were discovered and utilized in a variety of socially impactful electronic, optoelectronic, and memory devices, the second involves the purposeful manipulation of coherent quantum mechanical properties to realize novel and powerful technologies. There are several quantum-coherence based platforms to implement the second quantum revolution, including ion traps, Rydberg atoms in optical lattices, superconducting quantum interference devices and fully optical systems. Technologies utilizing quantum states of light in the form of single-photon states, indistinguishable photons, and entangled photon pairs are robust against decoherence due to low environmental interaction~\cite{gisinQuantumCryptography2002} and continue to play a major role either independently, or in conjunction with other platforms.
	
	%Though numerous solid-state, atomic, and nonlinear optical sources are being explored to hopefully one day meet said demands, semiconductor quantum dots (QDs) obtained by epitaxial growth are particularly promising candidates.
	
	QDs which can act as a triggered source of entangled photon pairs would be of particular use to a number of quantum computation, communication, and sensing schemes.~\cite{aharonovichSolidstateSinglephotonEmitters2016, orieuxSemiconductorDevicesEntangled2017, huberSemiconductorQuantumDots2018, senellartHighperformanceSemiconductorQuantumdot2017} Entangled photons from a QD are typically realized using the original proposal of \citeauthor{bensonRegulatedEntangledPhotons2000},~\cite{bensonRegulatedEntangledPhotons2000} whereby polarization-entangled photon pairs are produced via the biexciton-exciton cascade within a single, excited QD. Perfect entanglement of these photon pairs relies crucially on the energy degeneracy of the intermediate exciton levels in the radiative cascade, which in turn is heavily reliant on the symmetry of the electron-hole exchange interaction within the QD.~\cite{orieuxSemiconductorDevicesEntangled2017, huberSemiconductorQuantumDots2018} The exchange interaction becomes anisotropic when the QD is no longer rotationally invariant in the plane of growth,~\cite{bayerFineStructureNeutral2002} and when nonsymmetric strain and compositional gradients are present,~\cite{huberSemiconductorQuantumDots2018} resulting in a fine structure splitting (FSS) of the neutral exciton energy. Furthermore, the electron spin coherence is reduced in quantum dots with an inherent strain distribution;\cite{stockillQuantumDotSpin2016} it is therefore desirable, from the perspective of developing solid-state QD devices, to grow symmetric, unstrained QDs with well-defined three-dimensional confinement.
	
	QDs have historically been grown in the molecular beam epitaxy (MBE) and metal-organic vapor-phase epitaxy (MOVPE) environments by Stranski-Krastonov (SK) growth. In SK growth, strain is partially relaxed by three-dimensional island formation, with the typical material system for QD growth being In(Ga)As embedded in a GaAs matrix. Though this material was used almost exclusively in the pioneering works of QD experimentation, including early experimental realizations of entanglement from the biexciton-exciton cascade,~\cite{youngImprovedFidelityTriggered2006, akopianEntangledPhotonPairs2006} nearly all SK QDs suffer from reduced symmetry in the electron-hole exchange interaction. Furthermore, SK growth is relatively inflexible when it comes to varying QD sizes (which in turn determines emission energies), as it is often linked to QD densities.~\cite{solomonEffectsMonolayerCoverage1995, solomonSubstrateTemperatureMonolayer1995}
	
	In the last two decades an alternative growth method known as droplet etching epitaxy (DEE), also commonly known as local droplet etching, has played an increasingly large role in QD growth. First introduced by \citeauthor{wangNanoholesFabricatedSelfassembled2007},~\cite{wangNanoholesFabricatedSelfassembled2007} DEE for QD growth is based on the controlled in-situ etching of an epitaxial film from group-III droplets in an environment of low group-V overpressure. The nanoholes, which result from etching a given host semiconductor film by the droplets, constitute a template that can be regrown and filled in with a lower bandgap semiconductor. The subsequent growth of a capping layer, usually of the same composition as the etched film, produces the QDs. With careful control of the MBE environment, the grower is able to systematically create nanoholes that exhibit near perfect in-plane symmetry. For example, using GaAs to fill in nanoholes created on an \ce{AlGaAs} layer, QDs can be formed which exhibit high planar symmetry and very low strain gradients and material intermixing. Said QDs therefore have greater electron-spin coherence~\cite{nguyenEnhancedElectronSpinCoherence2023} and much lower FSS than their SK-grown predecessors, and indeed have supplanted them in many entanglement experiments that exploit the biexciton-exciton cascade.~\cite{dasilvaGaAsQuantumDots2021} Furthermore, GaAs QDs grown by DEE (embedded in microcavities) are amongst the highest quality solid-state single-photon emitters in terms of single-photon purity, indistinguishability, and
	brightness.~\cite{senellartHighperformanceSemiconductorQuantumdot2017, dasilvaGaAsQuantumDots2021}
	
	The focus of this work is a detailed review on the knowledge of DEE as an epitaxial growth process for the realization of high quality QDs. It is an additional aim of this work to add nuance to many of the observations on DEE QD growth, and to open up questions and possible research pathways on the processes governing QD growth. It is our hope that by combining a comprehensive study of the existing literature with our own observations, this review can be used by those new to QD growth, and motivate new research to those already in the field. The structure of the review is as follows. Section~\ref{sec:sample_growth} gives the experimental conditions used to realize DEE QDs in our lab, and the growth conditions for sample data used in this work. In Section~\ref{sec:drop_nuc} we review the initial droplet deposition and nucleation step, which serves to determine the final nanohole, and thus QD, density. Knowledge on the subsequent etching step is reviewed in Section~\ref{sec:etch} with emphasis on thermodynamic and kinetic aspects. The final redeposition step is covered in Section~\ref{sec:redeposition}. We conclude this review in Section~\ref{sec:optics} with a brief look at some of the optical properties of DEE QDs in the context of the growth conditions.
	
	\section{Method of Crystal Growth} \label{sec:sample_growth}
	\begin{figure*}
		\includegraphics[width=\textwidth]{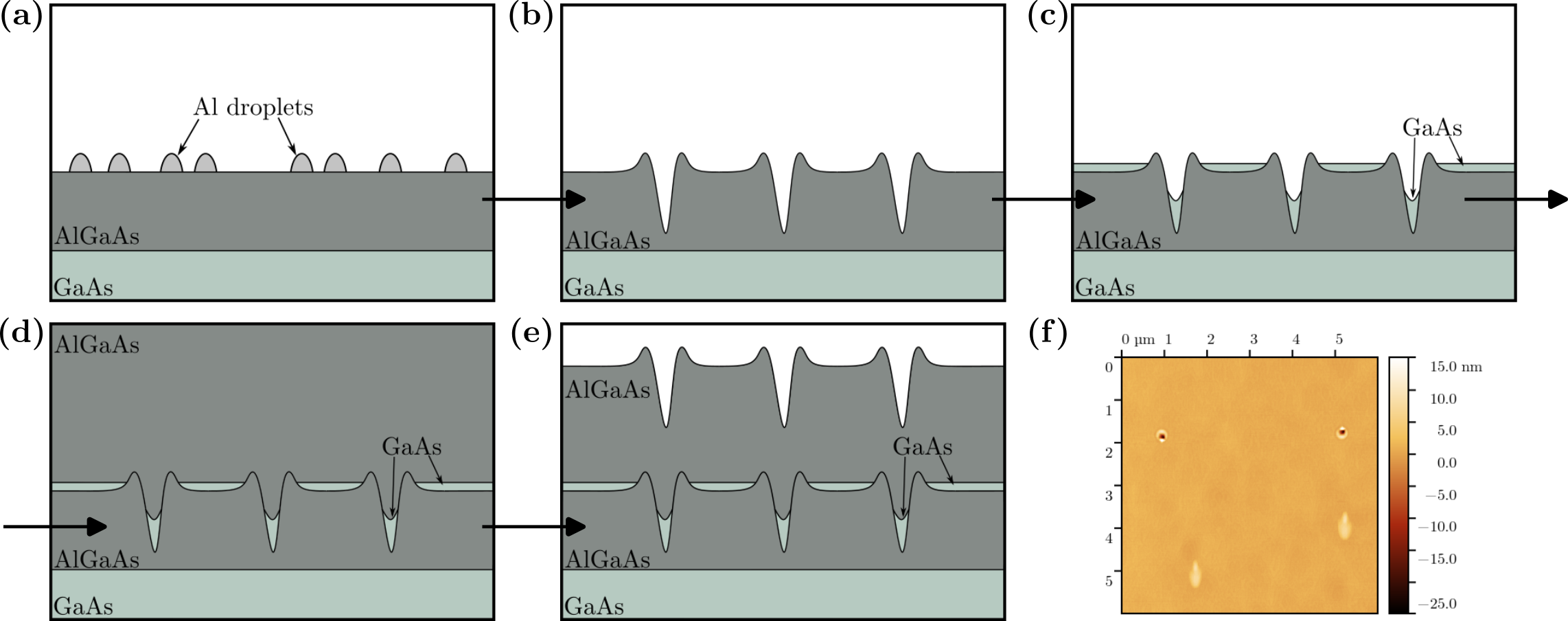}
		\caption{\textbf{(a)} - \textbf{(e)} Schematic of DEE for GaAs QD growth, demonstrating the principle growth steps.  \textbf{(a)} Growth begins with the deposition and nucleation of Al droplets on an AlGaAs buffer/confinement layer. \textbf{(b)} In the absence of a large As flux, the droplets etch into the buffer layer, forming nanoholes. \textbf{(c)} The nanoholes are regrown with GaAs. \textbf{(d)} QDs are functionalized by capping with AlGaAs. \textbf{(e)} The etching of nanoholes on the surface is an additional step taken for characterization of the growth. \textbf{(f)} An AFM image demonstrating buried QDs (raised elongated mounds) and nanoholes (ringed hole features).}\label{fig:droplet_deposition}
	\end{figure*}
	
	While results from other research groups are discussed, QD growths from the authors' group were completed in a Veeco GenXcel MBE system on 2 inch GaAs(001) substrates. The system was equipped with standard effusion cells for group-III sources, and a valved group-V source for As, with the cracker temperature at \qty{900}{\degreeCelsius} to produce a flux that is nearly completely cracked \ce{As_4}, i.e., \ce{As_2}. Oxide desorption under a constant As flux was monitored using reflection high-energy electron diffraction (RHEED) and used to calibrate our band-edge detection system (kSA BandiT), assuming desorption at \qty{580}{\degreeCelsius}.~\cite{springthorpeMeasurementGaAsSurface1987} After oxide removal, the substrate temperature was maintained at \qty{600\pm 5}{\degreeCelsius} for the entire growth. Sample growth began with a \qty{300}{\nano\meter} GaAs buffer layer followed by \qty{300}{\nano\metre} \ce{Al_{0.33}Ga_{0.67}As}. Aluminum was  deposited on the \ce{Al_{0.33}Ga_{0.67}As} surface at a flux, $F$ (with an error estimated at 2\% from RHEED oscillations) in the absence of an As overpressure, with the background arsenic beam equivalent pressure (BEP) measured to be between \qty{5.2E-9}{\torr} and \qty{8.0E-9}{\torr} immediately prior to deposition. Throughout deposition under these conditions, Al droplets are formed in the Volmer-Weber growth mode. An amount of Al equivalent to the Al in a \qty{1.2}{\monolayer} planar AlAs layer was deposited during the droplet nucleation step for all samples, with sample rotation maintained such that Al was homogeneously deposited across the substrate. After (and presumably during) Al deposition, liquid droplets form and begin to etch into the surface forming nanoholes, with the process allowed to run for \qty{180}{\second} in the absence of an As overpressure. The as-formed nanoholes were regrown with 1.5 to 7.5 equivalent planar monolyayers of GaAs, and a \qty{30}{\second} growth interrupt was applied for every \qty{1.5}{\monolayer} of deposited material to allow the GaAs to diffuse into the nanoholes. The regrown nanoholes were then capped with \qty{80}{\nano\metre} \ce{Al_{0.33}Ga_{0.67}As} to functionalize the GaAs QDs. Finally, Al deposition in the absence of an As overpressure  was repeated on the surface of the capping layer with the same parameters as earlier in the growth sequence. This produces nanoholes on the surface representative of the nanohole ensemble used to form the GaAs quantum dots,~\cite{kruckCriticalAluminumEtch2024} and permits topological characterization with atomic force microscopy (AFM). A schematic of the growths is given in Fig.~\ref{fig:droplet_deposition}.\\
	
		\begin{table*}
		\centering
		\begin{ruledtabular}
			\begin{tabular}{l|ccccc}
				Droplet[Epilayer] & $T$ (\unit{\degreeCelsius}) & $F$ (\unit{\monolayer\per\second}) & $F_\text{V}$ (BEP \unit{torr}) & $N$ (\unit{\per\micro\metre\squared}) & Source\\
				\hline
				Al[\ce{Al_{0.33}Ga_{0.67}As}] & 600 & \numrange{0.059}{1.25} & $< \num{1E-8}$ & \numrange{0.026}{0.7} & Fig. \ref{fig:flux_density}\\
				Al[\ce{Al_{0.35}Ga_{0.65}As}] & \numrange{570}{670} & 0.4 & $< \num{1E-7}$ & \numrange{0.04}{1.5} & Ref.~\onlinecite{heynScalingStructuralCharacteristics2014}\\
				Al[\ce{Al_{0.33}Ga_{0.67}As}] & 620 & 0.4 & \numrange{0.5E-7}{9E-7} & \numrange{0.03}{1.1}& Ref.~\onlinecite{sonnenbergHighlyVersatileUltralow2012}\\
				Ga[GaAs] & 520 & \numrange{0.05}{1} &  $< \num{1E-7}$\footnote{\label{firstfootnote}The authors state that the As shutter is closed during deposition, but do not give a value of the background overpressure. For a typical III-V MBE system this is probably consistent with a background overpressure $< \qty{1E-7}{\torr}$.}& \numrange{0.2}{2.5} & Ref.~\onlinecite{atkinsonIndependentWavelengthDensity2012}\\
				Ga[GaAs] & \numrange{480}{620} & 0.8 & $< \num{1E-7}$\footref{firstfootnote} & \numrange{2}{10} & Ref.~\onlinecite{heynKineticModelLocal2011}\\
				Ga[\ce{In_{0.4}Ga_{0.6}As}(111)A] & 540 & \numrange{0.3}{1} & \num{1.0E-7} & \numrange{5.5}{12.5} & Ref.~\onlinecite{tuktamyshevLocalDropletEtching2024}\\
				Al[AlAs] & \numrange{560}{620} & 0.4 & $< \num{1E-7}$ & \numrange{2}{18} & Ref.~\onlinecite{kerbstDensityLimitsHigh2014}\\
				In[GaAs]& \numrange{400}{540} & 0.8 & $< \num{1E-7}$ & \numrange{0.05}{0.17} & Ref.~\onlinecite{stemmannLocalDropletEtching2008}\\
				Ga[\ce{Al_{0.3}Ga_{0.7}Sb}]& \numrange{270}{500}& 0.7 & \qty{0.060}{\monolayer\per\second} & \numrange{0.06}{20} & Ref.~\onlinecite{hilskaNanoholeEtchingAlGaSb2021}\\
				\ce{In_{0.52}Al_{0.48}}[\ce{In_{0.52}Al_{0.48}As}]& \numrange{390}{505} & 0.68 & \num{1.35E-6} & \numrange{4.7E-3}{0.1} & Ref.~\onlinecite{deutschTelecomCbandPhoton2023}\\
			\end{tabular}
		\end{ruledtabular}
		\caption{Nanohole/droplet densities for a variety of DEE systems. In the table, $T$ is the substrate temperature at the time of deposition, $F$ is the flux of the group-III species to the epilayer, $F_\text{V}$ is the group-V overpressure during droplet formation, and $N$ is the experimentally observed nanohole density. Unless indicated otherwise, the surfaces are always (001). This table is not an exhaustive account of all published experimental nanohole densities observed in DEE. It intends to represent nanohole densities in key material systems with sufficient information published on the growth conditions. Values of $N$ are typically only given in figures, and thus the values tabulated above are estimates from the relevant text.}\label{tab:drop_den}
	\end{table*}

\section{Droplet Nucleation} \label{sec:drop_nuc}
Droplet etching epitaxy begins with the deposition of the group-III atomic species on a III-V epilayer in conditions of limited group-V overpressure, which is realized by a valved group-V source, reducing the group-V flux to the background group-V partial pressure remaining from earlier growth. The group-III adatoms are consumed by the surface reconstruction until the surface is saturated, whereupon nanoscale droplets form in a surface diffusion process from a uniform group-III deposition in the Volmer-Weber growth mode. If the stochiometry of the surface reconstruction is known, then the `critical' amount of group-III material consumed before the droplets can be formed is predictable.~\cite{heynRoleArsenicAluminum2016, hilskaNanoholeEtchingAlGaSb2021} In the absence of Ostwald ripening (see Sec.~\ref{sec:temp_dep}), and assuming every stable droplet etches a nanohole, one may take the nanohole density to be equal to the density of droplets on the epilayer at the end of the deposition step. In general, the droplet density is a function of the growth conditions, the most important of these being the deposition flux, substrate temperature, and group-V overpressure. Table~\ref{tab:drop_den} gives typical droplet densities for a number of DEE systems and research programs.

To model the densities and distributions of droplets in DEE, researchers have used a number of methods. Though Monte Carlo simulations appear an attractive option since they can model nucleation on the epilayer purely from energetic parameters (see Ref.~\onlinecite{ratschNucleationTheoryEarly2003} and references therein), the higher temperatures used in DEE and the large adatom fields required to model the relatively low droplet densities incur large computational costs.~\cite{heynModelingGaDroplet2021} Atomistic rate equations, which were largely developed by Venables,~\cite{venablesRateEquationApproaches1973, venablesNucleationGrowthThin1984} are a popular alternative for modelling droplet densities since their mean-field nature predicts analytic scaling laws for a number of limiting cases (or regimes) in nucleation from the vapor. The rate equations are consistent with Monte Carlo simulations for mean quantities such as average island size, adatom density, and total island density,\cite{balesDynamicsIrreversibleIsland1994} however they are completely inadequate for describing the distribution of island sizes due to the equation's mean-field nature. Because the structural features of the nanoholes can be directly related to the geometrical features of the droplets from which they are etched,~\cite{heynKineticModelLocal2011, liOriginNanoholeFormation2014, heynScalingStructuralCharacteristics2014, tuktamyshevLocalDropletEtching2024} the island size distribution is of some interest. In recent years a phenomenological approach to the island size distribution has been developed by Pimpinelli and Einstein, which proposes that the distribution in size of Voronoi polygons built around the observable stable clusters can be modelled by the generalized Wigner surmise (GWS).~\cite{pimpinelliCaptureZoneScalingIsland2007,pimpinelliPimpinelliEinsteinReply2010,pimpinelliScalingExponentEqualities2014} The application of the rate equations and the GWS to modelling the droplet nucleation step in DEE are discussed in the following.
	
\subsection{Rate equation approach to DEE}\label{sec:rate_eq}
	According to the mean-field approach, the areal density of stable clusters, $N$, during nucleation is given by\cite{venablesNucleationGrowthThin1984, venablesAtomicProcessesCrystal1994}
	\begin{equation}
		N \sim F^\alpha\exp\left(\frac{E}{k_B T}\right) \label{eq:scaling_law}
	\end{equation}
	where $F$ is the flux from the vapor, $k_B$ the Boltzmann constant, $E$ the nucleation activation energy, and $\alpha$ a scaling exponent. The derivation of Eq.~(\ref{eq:scaling_law}) requires the concept of the `critical nucleus' size $i$ from classical nucleation theory. In short, $i$ is defined by $i+1$ being the smallest cluster of atoms necessary to form a stable nucleus on the surface. Clusters of size $> i$ capture adatoms before they (on average) decay, whereas the opposite is true for subcritical clusters with size $< i$.  Furthermore, the population of subcritical clusters is assumed to be in local equilibrium with the vapor,\cite{venablesIntroductionSurfaceThin2000} and is accounted for in the rate equations by the Walton relation.~\cite{waltonNucleationVaporDeposits1962} Three different nucleation regimes occur in practice, the prevalence of any one being dependent upon the ratio $\tau_a / \tau_c$, where $\tau_a$ is the characteristic time for desorption of adatoms from the surface, and $\tau_c$ is the characteristic time for capture of an adatom by a stable cluster. Extreme incomplete condensation occurs when $\tau_a / \tau_c \ll 1$, i.e, when there is strong re-evaporation of the adatoms from the surface. In this regime, clusters can only grow by direct impingement of the atomic species from the flux. Complete condensation occurs when $\tau_a / \tau_c \gg 1$, which is realized in practice by systems where there is no re-evaporation of the adatoms. In between these two regimes is initially incomplete condensation in which $\tau_a / \tau_c \lesssim 1$ and where clusters grow by the surface diffusion of adatoms, but reevaporation of adatoms before capture into a stable cluster remains likely. The parameter dependence in Eq.~(\ref{eq:scaling_law}) for these regimes can be directly extracted from the rate equations,~\cite{venablesNucleationGrowthThin1984} or for $\alpha$ alone from scaling arguments.~\cite{pimpinelliScalingCrossoversModels1999} The results for 2D and 3D clusters are listed in Table~\ref{tab:par_dep}.
	
	\subsubsection{Flux dependence}
\begin{figure}
		\centering
		\includegraphics[width=\columnwidth]{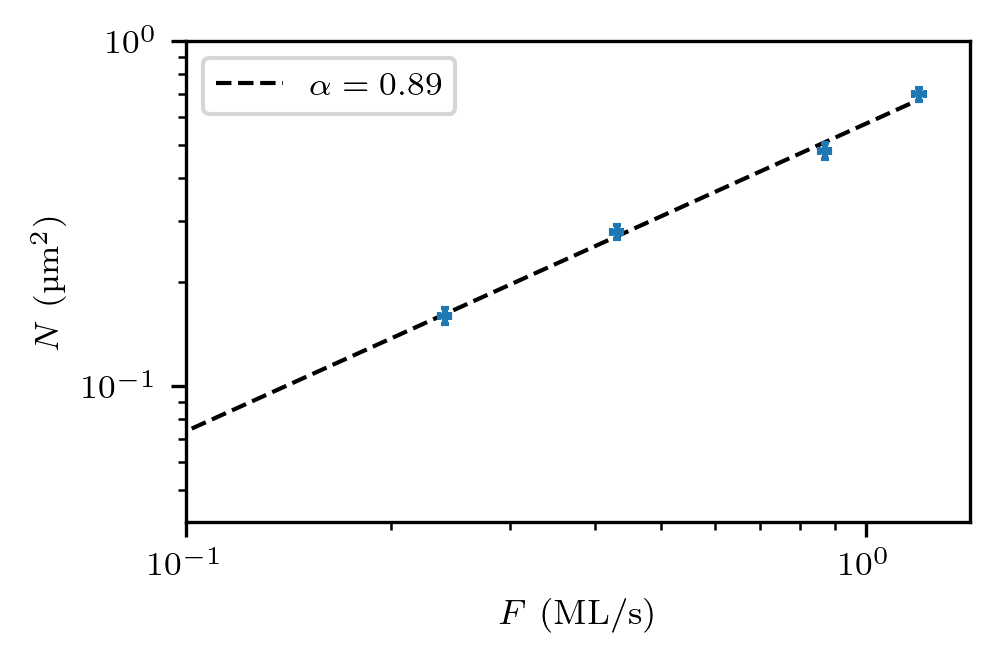}
		\caption{Density of nanoholes as a function of the Al deposition flux on \ce{Al_{0.33}Ga_{0.67}As}. The substrate temperature was \qty{600}{\degreeCelsius}, with the valve and shutter to the As cracker closed, resulting in $F_{\text{\ce{As_2}}} < \qty{1E-8}{\torr}$. The total deposited amount of Al was equivalent to \qty{1.2}{\monolayer} AlAs. Up to nine AFM scans were taken for each growth, with the 68\% confidence interval used to estimate the error in $N$. Nucleation is assumed to occur under complete condensation conditions since there exists no obvious process removing Al from the system. The least squares fit used in the figure gives $\alpha = 0.89 \pm 0.04$, which corresponds to $i = 24 \pm 10$.}\label{fig:flux_density}
	\end{figure}

	The nucleation of either Ga or Al droplets on the \ce{AlGaAs} surface is typically assumed to occur in the complete condensation regime because the substrate temperatures used in DEE are usually below that in which significant desorption of Ga,\cite{keanGalliumDesorptionAlGaAs1991} or Al, are known to occur. Values of the critical nucleus size $i$ can therefore be calculated by measuring $\alpha$ from a plot of $n_x$ versus $F$ and using the appropriate relation in Table~\ref{tab:par_dep}. Here we mention some key results for the values of $i$ found for Ga/Al droplets on the \ce{Al_xGa_{1-x}As} surface using Eq.~(\ref{eq:scaling_law}); Table \ref{tab:i_values} gives a comprehensive listing of $i$ from the literature. For Ga droplets on GaAs(001), \citeauthor{ohtakeExtremelyHighLowDensity2015}\cite{ohtakeExtremelyHighLowDensity2015} found $i = 5$ (no error given) at $T = \qty{200}{\degreeCelsius}$, whereas \citeauthor{atkinsonIndependentWavelengthDensity2012}\cite{atkinsonIndependentWavelengthDensity2012} found  $i = 13\pm 5$ at $T = \qty{550}{\degreeCelsius}$. \citeauthor{heynModelingGaDroplet2021}\cite{heynModelingGaDroplet2021} numerically integrated the rate equations instead of resorting to Eq.~(\ref{eq:scaling_law}), for Al droplets on \ce{Al_{0.35}Ga_{0.65}As}, and found from fits to the experimental nanohole densities a dependence of the critical nucleus size on temperature of the form $i \propto \exp(T)$. In particular, at $T = \qty{600}{\degreeCelsius}$, $i \approx 20$, whereas at $T=\qty{450}{\degreeCelsius}$, $i \approx 10$. This is not in agreement with $i = 5$ found by \citeauthor{wangMechanismAluminumDroplet2023}\cite{wangMechanismAluminumDroplet2023} for Al droplets on GaAs(001) at $T = \qty{480}{\degreeCelsius}$, however, this latter work only uses three data points, and gives no error in its value of $i$. Figure~\ref{fig:flux_density} presents results from our own lab on the nanohole density from Al droplet etching on \ce{Al_{0.33}Ga_{0.67}As} at \qty{600}{\degreeCelsius}. Assuming complete condensation conditions, we find $i = 24 \pm 10$, in agreement with \citeauthor{heynModelingGaDroplet2021}.~\cite{heynModelingGaDroplet2021}
	
	\begin{table}
		\centering
		\begin{ruledtabular}
			\begin{tabular}{l|ccc}
				Condensation regime & {} & $\alpha$ & $E$\\
				\hline
				\multirow{2}{*}{Extreme incomplete} & 2D\footnote{For extreme incomplete condensation of 2D clusters, \citeauthor{pimpinelliScalingCrossoversModels1999}\cite{pimpinelliScalingCrossoversModels1999} give $\alpha=2i/3$ whereas \citeauthor{venablesNucleationGrowthThin1984}\cite{venablesNucleationGrowthThin1984} give  $\alpha = i$. The former result is taken for reasons given in Ref.~\onlinecite{jensenEffectMonomerEvaporation1997}, with the value of $E$ adjusted accordingly.} & $2i/3$ & $(2/3)(E_i + (i+1)E_a - E_d)$ \\
				& 3D  & $2i/3$  & $(2/3)(E_i + (i+1)E_a - E_d)$ \\
				\midrule
				\multirow{2}{*}{Initially incomplete} & 2D & $i/2$ & $(1/2)(E_i +iE_a)$ \\
				& 3D  & $2i/5$  & $(2/5)(E_i + iE_a)$\\
				\midrule
				\multirow{2}{*}{Complete} & 2D &  $i/(i+2)$ & $(E_i + iE_d)/(i+2)$\\
				& 3D   &  $i/(i + 2.5)$  & $(E_i + iE_d)/(i+2.5)$\\
			\end{tabular}
		\end{ruledtabular}
		\caption{Parameter dependence for the stable cluster density in Eq.~(\ref{eq:scaling_law}), with $E$ the nucleation activation energy, $\alpha$ the scaling exponent, and $i$ the critical nucleus size. In general, $E$ is a function of the activation energies for the atomistic processes modelled in the rate equations. These include the binding energy of a critically sized cluster $E_i$, the re-evaporation energy for an adatom $E_a$, and the surface diffusion energy for an adatom $E_d$.}\label{tab:par_dep}
	\end{table}
	
	\begin{table*}
		\begin{ruledtabular}
		\centering
		\begin{tabular}{l|cccc}
				\toprule
				Droplet[Epilayer] & $T$ (\unit{\degreeCelsius}) &  $i$ & Method & Source\\
				\midrule\midrule
				Ga[GaAs] & 200 & 5 & Scaling & Ref.~\onlinecite{ohtakeExtremelyHighLowDensity2015}\\
				Ga[GaAs] & 200 & 3 & Scaling & Ref.~\onlinecite{heynRegimesGaAsQuantum2007}\\
				Ga[GaAs] & \numrange{185}{200} & $5 \pm 1$ to $7 \pm 1$ & GWS & Ref.~\onlinecite{nothernTemplatedependentNucleationMetallic2012}\\
				Ga[GaAs] & 520 & $13 \pm 5$ & Scaling & Ref.~\onlinecite{atkinsonIndependentWavelengthDensity2012}\\
				Ga[GaAs(111)A] & 200 & 1 & Scaling & Ref.~\onlinecite{ohtakeExtremelyHighLowDensity2015} \\
				Ga[GaAs(111)A]\footnote{Growth surface is vicinal with a miscut of \qty{2}{\degree} towards $(\bar{1}\bar{1}2)$.} & \numrange{300}{450} & $3 \pm 1$ and $2\pm 0.5$\footnote{For $T < \qty{400}{\degreeCelsius}$, \citeauthor{tuktamyshevNucleationGaDroplets2021} give $i=3\pm 1$. For $T > \qty{400}{\degreeCelsius}$, the authors give $i = 2\pm0.5$ due to a change in the diffusion conditions associated with the vicinal epilayer.} & GWS & Ref.~\onlinecite{tuktamyshevTemperatureActivatedDimensionality2019}\\
				Ga[GaAs(111)A] & \numrange{300}{450} & $1.0 \pm 0.4$ to $4.0 \pm 0.3$ & GWS & Ref.~\onlinecite{tuktamyshevNucleationGaDroplets2021}\\
				Ga[\ce{In_{0.4}Ga_{0.6}As}(111)A] & 540 & $5.4 \pm 1.5$\footnote{\citeauthor{tuktamyshevLocalDropletEtching2024} give $i = 1.675 \pm 0.150$ consistent with initially incomplete condensation. The value tabulated here is for complete condensation; see text for details.} & Scaling & Ref.~\onlinecite{tuktamyshevLocalDropletEtching2024} \\
				Al[GaAs] & 480 & 5\footnote{\citeauthor{wangMechanismAluminumDroplet2023} give values of $i$ equal to 1, 2, and 5, depending on the nucleation regime. Consistent with other works we assume nucleation to be in the complete condensation regime for this system.} & Scaling & Ref.~\onlinecite{wangMechanismAluminumDroplet2023} \\
				Al[\ce{Al_{0.4}Ga_{0.6}As}] & 600 & 2 & GWS & Ref.~\onlinecite{loblCorrelationsOpticalProperties2019}\\
				Al[\ce{Al_{0.33}Ga_{0.67}As}] & 600 & $24 \pm 10$ & Scaling & Fig. \ref{fig:flux_density}\\
				Al[\ce{Al_{0.33}Ga_{0.67}As}] & 600 & $4.0 \pm 0.2$ & GWS + Scaling & Fig. \ref{fig:droplet_tesselation}(b)\\
				\bottomrule
		\end{tabular}
		\end{ruledtabular}
		\caption{Experimentally determined values of the critical nucleus size $i$ for material systems relevant to DEE. Unless otherwise given, all surfaces are (001).}\label{tab:i_values}
	\end{table*}
	
	\subsubsection{Effect of a group-V overpressure}
	\citeauthor{tuktamyshevLocalDropletEtching2024}\cite{tuktamyshevLocalDropletEtching2024} studied nanoholes etched by Ga droplets on \ce{In_{0.4}Ga_{0.6}As}(111)A at \qty{540}{\degreeCelsius} and argued that a channel exists for adatom removal during deposition via reaction with the background As. In effect, the characteristic time for desorption $\tau_a$ can instead be used to describe the characteristic time for reaction of the adatoms with the background As on the surface. As such, $\tau_a$ would decrease as the As overpressure increases and the ratio $\tau_a / \tau_c$ would cross over into the initially incomplete condensation regime. \citeauthor{tuktamyshevLocalDropletEtching2024}\cite{tuktamyshevLocalDropletEtching2024} state that for the As overpressure used during growths (\ce{As_4} BEP of \qty{1.0E-7}{\torr}), droplet nucleation occurs under conditions of initially incomplete condensation, with $i = 1.675 \pm 0.150$ physically equivalent to $i=2$. However, we note that the As overpressure used in their work is no higher than is typical of other DEE growths, and retaining the assumption of complete consumption gives  $i = 5.4 \pm 1.5$. This would be consistent with previous works from the same group for Ga droplets on GaAs(111)A surfaces, where it was shown that $i = 5 \pm 1$ at \qtyrange{300}{350}{\degreeCelsius}\cite{tuktamyshevTemperatureActivatedDimensionality2019} and $i = 4 \pm 0.3$ at \qty{450}{\degreeCelsius}.~\cite{tuktamyshevNucleationGaDroplets2021}
	
	Nonetheless, the analysis does suggest that for a large enough group-V overpressure during deposition (short of group-III rich epitaxial growth), the condensation regime can change to initially incomplete. To our knowledge, only the works of \citeauthor{sonnenbergHighlyVersatileUltralow2012}\cite{sonnenbergHighlyVersatileUltralow2012} and \citeauthor{deutschLocalDropletEtching2025}\cite{deutschLocalDropletEtching2025} have systematically investigated the effect of group-V overpressure on droplet density. In both works, interpretation of the data is complicated by the existence of a bimodal distribution in the droplet sizes. Thus, the sharp change from a bimodal to unimodal nanohole distribution with increasing As overpressure may indicate a transition in nucleation regime. Alternatively, droplets will be consumed quicker with an increase in As overpressure~\cite{heynKineticModelLocal2011, liOriginNanoholeFormation2014, fusterFundamentalRoleArsenic2014} (regardless of nucleation regime) and this may produce a range in overpressures whereby small droplets are consumed before they can measurably etch the epilayer, whereas larger droplets continue to produce nanoholes. AFM images from other works~\cite{heynInfluenceGaCoverage2009, heynRoleArsenicAluminum2016} that demonstrate nanoholes at variable As overpressures would appear to support a sharp change in nanohole densities, however quantitative information on the densities from these works are lacking, and so the problem remains open.

	\subsubsection{Temperature dependence and Ostwald ripening}\label{sec:temp_dep}
	From Eq.~(\ref{eq:scaling_law}), the temperature dependence of $N$ (at constant $F$) is expected to follow an Arrhenius relation, with densities generally decreasing with increasing $T$ (see Table~\ref{tab:drop_den}). Activation energies $E$ have been given for a number of systems in DEE from temperature dependent analyses of the droplet density,~\cite{heynKineticModelLocal2011, kerbstDensityLimitsHigh2014, heynScalingStructuralCharacteristics2014, hilskaNanoholeEtchingAlGaSb2021, tuktamyshevNucleationGaDroplets2021, tuktamyshevLocalDropletEtching2024} however, it is important to note that a number of complicating factors limit the usefulness of these measurements. For instance, abrupt changes in $E$ can occur due to changes in the surface reconstruction with growth temperature.~\cite{kerbstDensityLimitsHigh2014, hilskaNanoholeEtchingAlGaSb2021, tuktamyshevLocalDropletEtching2024} In this case, the change in $E$ is due to a change in the energy for surface diffusion $E_d$ of adatoms (see Table~\ref{tab:par_dep}), which have been given in a few instances for systems relevant to DEE,~\cite{biettiGalliumSurfaceDiffusion2014, stevensSurfaceDiffusionMeasurements2017, hilskaNanoholeEtchingAlGaSb2021, wangMechanismAluminumDroplet2023} but can itself have a complex dependence on temperature outside of surface reconstruction.~\cite{venablesNucleationGrowthThin1984, heynModelingGaDroplet2021} One such example is the abrupt change in $E$, with growth temperature, due to a change in the dimensionality of adatom diffusion on a vicinal epilayer.~\cite{tuktamyshevTemperatureActivatedDimensionality2019} The critical nucleus size $i$ also increases with temperature,\cite{venablesNucleationGrowthThin1984, jonesEnergiesControllingNucleation1990, heynModelingGaDroplet2021} and is a trend which can be observed in Table~\ref{tab:i_values}. This originates from the temperature activated detachment of adatoms from clusters, and as is evident from Table~\ref{tab:par_dep}, an increase in $i$ and $E_i$ (due to a larger number of pairwise bonds for atoms in the cluster), increases $E$. Thus, an Arrhenius relation for $N$ is only valid in limited temperature regions where $i$ may be taken as constant.
	
	The effect of coarsening due to Ostwald ripening of the droplets (defined here as the minimization of the total droplet free energy via the consumption of smaller droplets by larger droplets) must also be considered in an analysis of droplet nucleation. From a kinetics perspective, droplet ripening can occur from the dissolution of smaller droplets, but also from droplet motion,\cite{tersoffRunningDropletsGallium2009, kanjanachuchaiSelfRunningGaDroplets2013} which can be observed near the congruent evaporation temperature of the epilayer as trenches on the surface. An extended scaling law for the evolution of the droplet density as a function of time $t_r$, under conditions of Ostwald ripening,\cite{zinke-allmangClusteringSurfaces1992} was first given by \citeauthor{heynRegimesGaAsQuantum2007}.~\cite{heynRegimesGaAsQuantum2007} The scaling law is
	\begin{equation}\label{eq:extend_scaling_law}
		N(t_r) = N(0)(1 + t_r / \tau_r)^{-1}
	\end{equation}
	where $N(0)$ is identified with the density as given in Eq.~(\ref{eq:scaling_law}), and $\tau_r$ is assumed to have a temperature activated dependence of the form $\tau_r = \nu^{-1}\exp{(E_r / k_B T)}$, where $\nu = 2k_B T/ h$ is a vibrational frequency\cite{heynDynamicsSelfassembledDroplet2009} and $h$ is the Planck constant. Ostwald ripening occurs in systems where, at constant deposition flux, there is a reduction in droplet density as a function of total deposited droplet material.~\cite{heynNanoholeFormationAlGaAs2009,  nothernTemplatedependentNucleationMetallic2012,tuktamyshevLocalDropletEtching2024} In such cases, the increase in total droplet material is equivalent to an increase in deposition time. Coarsening is also observed in systems where the droplet density reduces as a function of the time given to deposition, or post deposition annealing,~\cite{heynRegimesGaAsQuantum2007, heynDynamicsSelfassembledDroplet2009, heynScalingStructuralCharacteristics2014}  and can occur concomitantly with etching.~\cite{heynInfluenceGaCoverage2009, heynDynamicsSelfassembledDroplet2009}
	
	Ostwald ripening has been used to explain the density of Ga\cite{heynRegimesGaAsQuantum2007, heynDynamicsSelfassembledDroplet2009, heynKineticModelLocal2011, ohtakeExtremelyHighLowDensity2015} and Al\cite{heynScalingStructuralCharacteristics2014, zocherAlloyingLocalDroplet2019, wangMechanismAluminumDroplet2023} droplets which do not follow an Arrhenius relation. However, an analysis of this is once again complicated by the fact that the critical nucleus size $i$ increases as a function of epilayer temperature. Thus, a change in $E$ as a function of temperature may well be associated with a change in $i$ as opposed to ripening, and it is for this reason that \citeauthor{heynModelingGaDroplet2021}\cite{heynModelingGaDroplet2021} revised their earlier conclusion that Ostwald ripening was a relevant process in Al droplet nucleation.~\cite{zocherAlloyingLocalDroplet2019} The extent of ripening in DEE systems requires further investigation, and can probably only be reliably decoupled from changes in $i$ for growths at constant temperature with very low As overpressure (so that droplet consumption is slow) and a time resolved annealing step.~\cite{heynDynamicsSelfassembledDroplet2009, fusterFundamentalRoleArsenic2014}
	
	\subsubsection{Bimodal size distributions}
	Bimodal size distributions in the nanoholes etched from droplets are frequently observed in DEE, typically appearing as populations of ``shallow'' nanoholes with small openings on the surface and ``deep'' nanoholes with much larger openings.~\cite{heynNanoholeFormationAlGaAs2009, sonnenbergHighlyVersatileUltralow2012, heynScalingStructuralCharacteristics2014, heynRoleArsenicAluminum2016, kusterDropletEtchingDeep2016, deutschLocalDropletEtching2025,massonEngineeringNanoholeEtchedQuantum2026} For Al droplet etching, a bimodal size distribution is seen to occur at high substrate temperature,~\cite{kusterDropletEtchingDeep2016} or high monolayer coverage of droplet material,~\cite{chelluHighlyUniformGaSb2021,massonEngineeringNanoholeEtchedQuantum2026} whereas the opposite is apparently true for Ga droplet etching.~\cite{heynNanoholeFormationAlGaAs2009} In other cases, the bimodal distribution is not so obvious, and may present only as two populations with different radii of the nanohole openings,~\cite{alonso-gonzalezLowDensityInAs2008} or in the height/size of the droplets.~\cite{lyamkinaInvestigationIntermediateStage2010} Barring pathological cases where nanoholes form at the sites of surface defects,~\cite{heynDynamicsMassTransport2015, cesuraDropletFreeSelfassembling2024} it is not generally well understood why a bimodal distribution appears. In the case of Ga droplet etching, the distribution in nanohole sizes appears to be consistent with droplet coarsening by Ostwald ripening and depends on the interplay between the time afforded for ripening, and the time required to etch the nanoholes.~\cite{heynNanoholeFormationAlGaAs2009} Thus in the coarsening process one has populations of small droplets feeding larger droplets; these smaller droplets continue to etch into the epilayer and are quickly consumed, producing a population of shallow nanoholes, whereas the larger droplets may persist on the surface for some time, eventually resulting in a population of deep nanoholes.~\cite{heynDynamicsSelfassembledDroplet2009, heynInfluenceGaCoverage2009, raabOswaldRipeningShape2000}
	
	Other surface phenomena may also play a role in the formation of bimodal droplet distributions, particularly in Al droplet etching systems. We suggest two relevant processes here in the interest of stimulating further research. Firstly, surface reconstruction changes occur during droplet formation,\cite{lyamkinaInvestigationIntermediateStage2010, heynScalingStructuralCharacteristics2014, heynRoleArsenicAluminum2016} and may significantly modify the diffusivity of the group-III adatoms and their subsequent nucleation centers.~\cite{kruckCriticalAluminumEtch2024} Secondly, surface segregation effects, in particular the segregation of Ga from the \ce{AlGaAs} growth front during AlAs growth,~\cite{jusserandLongRangeGallium1992, braunSituTechniqueMeasuring1994} will give rise to two adatom populations (Al and Ga) with different surface diffusivities. Ga surface segregation has previously been used by \citeauthor{zocherAlloyingLocalDroplet2019}~\cite{zocherAlloyingLocalDroplet2019} to explain the existence of liquid droplets during Al droplet etching of \ce{AlGaAs} surfaces at temperatures well below the Al melting point. Monte Carlo simulations predict bimodal island size distributions from such surface segregation under the right growth conditions,~\cite{mirandaBimodalIslandsizeDistributions2001} though distinct droplet populations have yet to be shown for Al droplet etching on \ce{AlGaAs}. We note, however, that \citeauthor{caoLocalDropletEtching2022}~\cite{caoLocalDropletEtching2022} have demonstrated distinct In and Al droplet populations for the etching of \ce{In_{0.55}Al_{0.45}As} with \ce{In_{0.55}Al_{0.45}} droplets. %The phsyical origins behind the onset of bimodal droplet and nanohole distributions in DEE remains an open research question 
	
\subsection{Capture zone approach}\label{sec:capture_zone}
To model the size distribution of stable clusters, statistical fluctuations in the local adatom density must be accounted for. Under certain conditions, the size distribution can be subordinated to the distribution in areas of the Voronoi polygons built around the clusters.~\cite{mulheranOriginsIslandSize1995,mulheranCaptureZonesScaling1996,blackmanScalingBehaviorSubmonolayer1996} Using such a scheme, a phenomenological approach taken by \citeauthor{pimpinelliCaptureZoneScalingIsland2007}~\cite{pimpinelliCaptureZoneScalingIsland2007, pimpinelliPimpinelliEinsteinReply2010} proposes that the areal distribution can be modelled by the generalized Wigner surmise (GWS) $P_\beta (s)$,
	\begin{equation}
		P_\beta (s) = a_\beta s^\beta \exp(-b_\beta s^2)\label{eq:GWS}
	\end{equation}
	where $s$ is the normalized area and $\beta$ is the sole parameter.\footnote{The constants in Eq.~\ref{eq:GWS} are derived from the conditions of unit probability and $\langle s \rangle = 1$. They are $\alpha_{\beta} = 2\Gamma\left(\frac{\beta+2}{2} \right)^{\beta+1} / \Gamma\left( \frac{\beta+1}{2} \right)^{\beta+2}$ and $b_{\beta} = \left[ \Gamma\left( \frac{\beta+2}{2} \right) / \Gamma\left( \frac{\beta+1}{2} \right)\right]^{2}$
	where $\Gamma$ is the Gamma function. } The parameter $\beta$ is related to the critical nucleus size by 
	\begin{equation}
		\beta = \gamma i + 1 + \gamma, \label{eq:beta}
	\end{equation} 
	where $\gamma$ is a characteristic of the limiting process in nucleation.~\cite{pimpinelliScalingExponentEqualities2014} For diffusion limited aggregation, which may be identified with the complete condensation regime in Table~\ref{tab:par_dep}, $\gamma = 1$ for 2D isotropic diffusion of the adatoms, increasing to $\gamma = 2$ for anisotropic 1D diffusion. For attachment limited aggregation, which can be tentatively identified with initially incomplete condensation, $\gamma = 1/2$, with more complicated dependencies possible. However, a far more powerful approach to measuring $i$ is given by the relation~\cite{pimpinelliScalingExponentEqualities2014}
	\begin{equation}
		\alpha (\beta + d/2 -1) = i,\label{eq:PE_relation}
	\end{equation}
	where $d$ is the dimensionality of the clusters. In the case of droplets on a surface, $d=3$. From Eq.~(\ref{eq:PE_relation}), $i$ can in principle be determined at constant temperature without knowledge of the nucleation regime or aggregation mechanism by fitting $\alpha$ from a plot of $N(F)$, and $\beta$ to the distribution of Voronoi areas in a sufficiently large scan of the surface.
	
	The GWS has in a few instances been applied to the nucleation of droplets in the GaAs and AlGaAs systems (see Table~\ref{tab:i_values} for such instances), however all cases assume irreversible nucleation and make use of the direct relation Eq.~(\ref{eq:beta}), rather than Eq.~(\ref{eq:PE_relation}). \citeauthor{nothernTemplatedependentNucleationMetallic2012}\cite{nothernTemplatedependentNucleationMetallic2012} found critical cluster sizes for Ga droplets deposited on GaAs(001) ranged from $i = 5 \pm 1$ at \qty{185}{\degreeCelsius} to $i = 7 \pm 1$ at \qty{200}{\degreeCelsius}. This is broadly in agreement with $i=5$ at \qty{200}{\degreeCelsius} found by \citeauthor{ohtakeExtremelyHighLowDensity2015}\cite{ohtakeExtremelyHighLowDensity2015} using the scaling relation Eq.~(\ref{eq:scaling_law}). \citeauthor{tuktamyshevNucleationGaDroplets2021}\cite{tuktamyshevNucleationGaDroplets2021} found $i = 2\pm1$ for temperatures $\leq \qty{400}{\degreeCelsius}$, increasing to $i=4 \text{ or } 5$ at \qty{450}{\degreeCelsius}, for Ga droplets on GaAs(111)A. 
	
	 \begin{figure}
		\centering
		\includegraphics[width=\columnwidth]{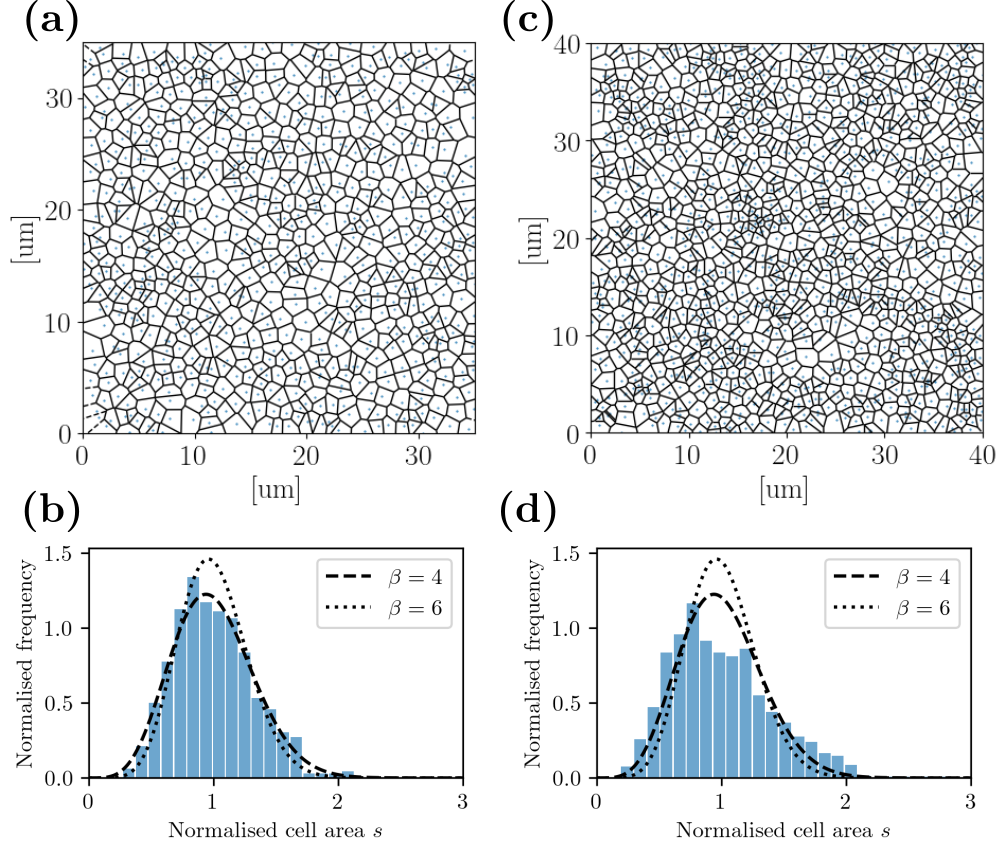}
		\caption{\textbf{(a)} and \textbf{(c)} Tessellation of two droplet etched surfaces. Blue dots in the tessellations correspond to the locations of nanoholes on the surface as determined by AFM scans. \textbf{(b)} and \textbf{(d)} Corresponding distribution of Voronoi cell areas. For \textbf{(a)} - \textbf{(b)} the growth parameters were $F = \qty{1.25}{\monolayer\per\second}$ and $T=\qty{600}{\degreeCelsius}$. In \textbf{(c)} and \textbf{(d)}, $F = \qty{0.28}{\monolayer\per\second}$ and $T=\qty{550}{\degreeCelsius}$. A generalized Wigner surmise with $\beta = 4$ fits the distribution in \textbf{(b)}, with $\beta = 6$ shown for comparison. In \textbf{(d)}, the unimodal GWS is not a suitable fit for what is clearly a bimodal distribution. The bimodal nature of the Voronoi cell area can be seen directly in \textbf{(c)}.}\label{fig:droplet_tesselation}
	\end{figure}
		
	The only example of the GWS being used in the DEE regime is by \citeauthor{loblCorrelationsOpticalProperties2019},\cite{loblCorrelationsOpticalProperties2019} who found, for Al droplets on \ce{Al_{0.4}Ga_{0.6}As} at \qty{600}{\degreeCelsius}, $i \approx 2$. This is contrary to the general trend in $i$ observed in Table~\ref{tab:i_values}, and to assess the validity of this result, we present in Fig.~\ref{fig:droplet_tesselation} Voronoi tessellations for two DEE growths done in our lab. In general, the low density of nanoholes in DEE systems can make the acquisition of suitable statistics for the GWS difficult. On our AFM system (Veeco Dimension 3100), scan sizes were limited to $\qty{80}{\micro\metre}\times\qty{80}{\micro\metre}$ and sufficient statistics could only be built up for growths with areal density $N \gtrsim \qty{1}{\per\micro\metre\squared}$. Figures~\ref{fig:droplet_tesselation}(a) - (b) show the Voronoi tessellation and GWS for DEE in which $F = \qty{1.25}{\monolayer\per\second}$ and $T = \qty{600}{\degreeCelsius}$ (rightmost data point in Fig.~\ref{fig:flux_density}). This growth is very similar to the one analyzed in the capture zone approach by \citeauthor{loblCorrelationsOpticalProperties2019},\cite{loblCorrelationsOpticalProperties2019} and indeed the distribution in $s$ is best fit with $\beta = 4$. Making use of the regime invariant Eq.~\ref{eq:PE_relation} and $\alpha  = 0.89 \pm 0.04$ gives $i = 4.0 \pm 0.2$, which is much closer to $i$ that is found assuming initially incomplete condensation in Fig.~\ref{fig:flux_density} ($i = 2.4 \pm 0.2$), rather than $i$ found from assuming complete condensation. In contrast, Figs.~\ref{fig:droplet_tesselation}(c) - (d) gives the tessellation and GWS for a growth in which $F = \qty{0.28}{\monolayer\per\second}$ and $T = \qty{550}{\degreeCelsius}$. Unlike the previous analysis, the distribution in $s$ is not well modelled by the GWS, demonstrating a bimodal distribution in the cell sizes. In fact, we observe a bimodal distribution in $s$ for most growths, with broadly unimodal distributions only apparent under conditions where the deposition flux $F$ is high. This is a clear indication of droplet coarsening as a function of deposition time,\cite{nothernTemplatedependentNucleationMetallic2012} and hence droplet sizes are not solely determined by their capture zones at the time of nucleation. Thus, even in the case of high deposition flux, where ripening times are expected to be low, coarsening effects cannot be disregarded, and fitting of the GWS to the cell areas cannot be reliably used to extract the critical nucleus size $i$, at least for Al droplet etching on \ce{AlGaAs}. %The existance of bimodal size distributions for Ga droplets on GaAs~\cite{alonso-gonzalezLowDensityInAs2008,lyamkinaInvestigationIntermediateStage2010} likewise suggests that GWS fitting cannot be applied to Ga droplet etching for the determination of the critical nucleus size.
	
	\section{Droplet Etching} \label{sec:etch}
	 At low group-V overpressure and high substrate temperature, the group-III droplets etch into epilayer, forming the nanohole templates that will later be regrown to produce QDs. The driving force behind the etching action is the realization of some equilibrium solubility of the group-V species in the group-III rich droplets.~\cite{wangNanoholesFabricatedSelfassembled2007} The droplets achieve this equilibrium solubility by liquifying the epilayer beneath them, thereby initiating etching. However, etching would quickly cease (when the equilibrium solubility is met) without a process removing the dissolved group-V species. Here, the surrounding epilayer and the rings observed around the nanoholes act as material sinks for the saturated droplet material, establishing a concentration gradient of the group-V species from the droplet-epilayer interface to the droplet periphery on the surface.~\cite{wangNanoholesFabricatedSelfassembled2007, stemmannLocalDropletEtching2008,heynInfluenceGaCoverage2009} The redistribution of the droplet material into the ring and epilayer is similar to the vapor-liquid-solid (VLS) growth mechanism familiar from nanowire growth. From an atomistic perspective, we may understand the three-phase boundary at the droplet-epilayer-vapor interface to have the locally highest concentration of the group-V species; having arrived directly from the background overpressure, from diffusion on the surface, and from within the droplet itself.~\cite{vasilenkoFormationGaAsNanostructures2015} This supersaturation of the group-V species at the three-phase boundary therefore precipitates out crystalline III-V material from the droplet material, forming the distinctive rings.
	 
	 Evidently, droplet etching can be understood as a combination of equilibrium thermodynamics and atomistic kinetics. Equilibrium thermodynamics dominates at high temperatures and is useful in predicting the limiting shape of the etch pits.~\cite{vonkFacetingLocalDropletetched2018} Furthermore, thermodynamics is the driving force behind preferential interface nucleation (i.e. VLS growth),\cite{wacaserPreferentialInterfaceNucleation2009} which can be used to explain the formation of rings around the nanoholes, and the redistribution of droplet material. Meanwhile, kinetics details the relevant reactions and exchanges that govern the droplet etching action. These kinetic descriptions once again come in the forms of Monte Carlo~\cite{reyesUnifiedModelDroplet2013, vasilenkoFormationGaAsNanostructures2015} and rate equation approaches.~\cite{heynKineticModelLocal2011, liOriginNanoholeFormation2014} Monte Carlo approaches to droplet etching are particularly useful for revealing the key reactions in droplet etching, and we summarize the main predictions of these simulations and the supporting evidence from growths. Some additional observations on droplet etching which do not neatly fit within a thermodynamics or kinetics description are given at the end of this section.
	
	\begin{figure*}
		\includegraphics[width=\textwidth]{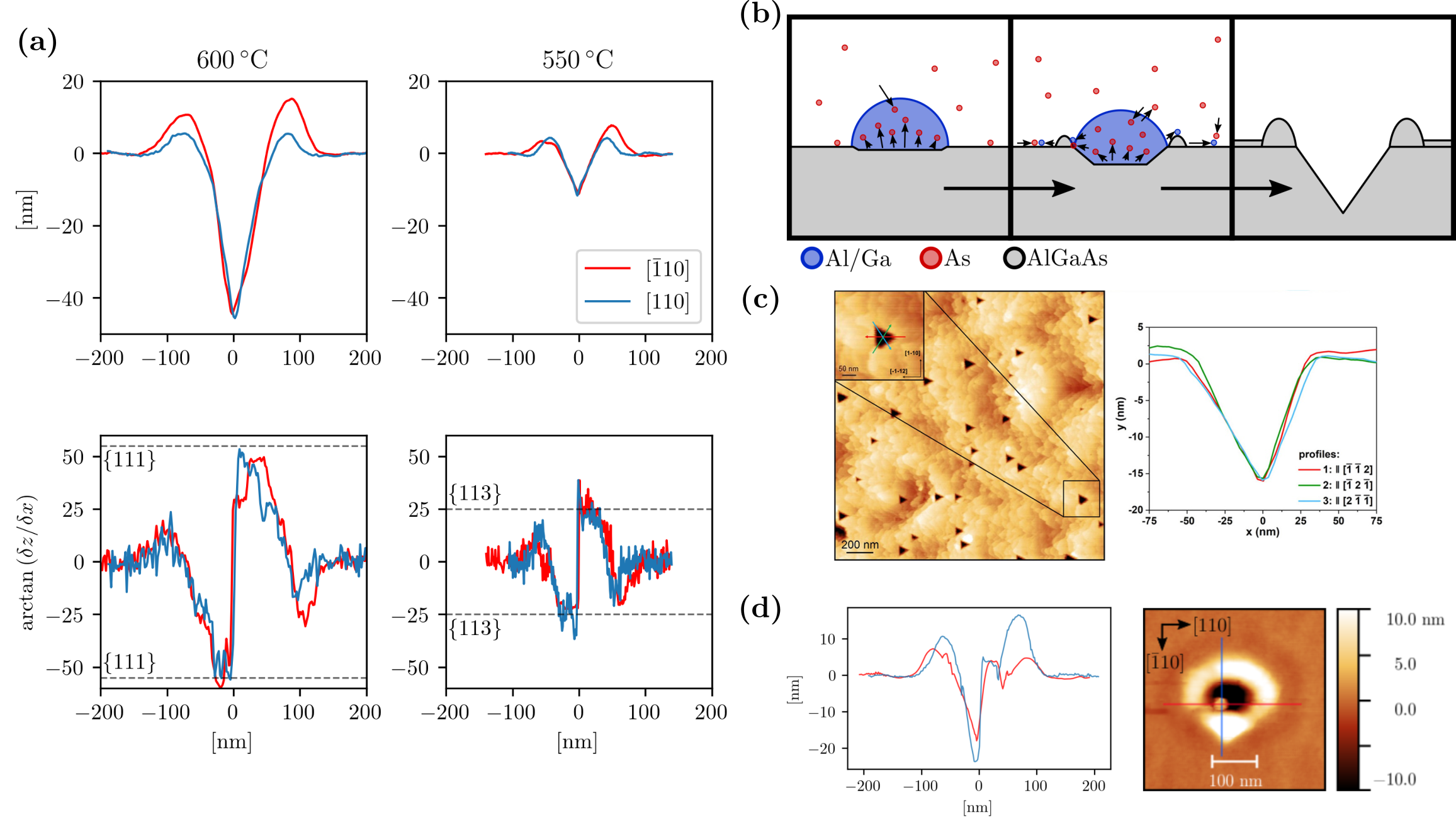}
		\caption{\textbf{(a)} Perpendicular linescans and derivative maps of two nanoholes, one etched at \qty{600}{\degreeCelsius} and the other at \qty{550}{\degreeCelsius}. For each growth, \qty{1.2}{\monolayer} of droplet material was deposited at a flux of \qty{0.25}{\monolayer\per\second} and under a background As BEP $< \qty{8E-8}{\torr}$. The slope/derivative of the linescans is calculated from $\delta z / \delta x$, where $z$ is perpendicular to the growth front, and $x$ parallel. The arctangent of the slope gives the angle of inclination to the $(001)$ surface. The $\{111\}$ and $\{113\}$ planes marked on the bottom row of graphs are inclined to the $(001)$ surface by approximately \qty{55}{\degree} and \qty{25}{\degree}, respectively. \textbf{(b)} Illustration of the evolution from droplets to nanoholes for Al/Ga droplet-etching on AlGaAs. As is first dissolved from the epilayer into the droplet, initiating etching. Droplet material is consumed in planar growth with the background As flux, and arsenides precipitate out at the three-phase boundary of the system forming a ring. The result is a concentration gradient of As within the droplet which allows for continued etching of the epilayer underneath the droplet. The final state of the system is a nanohole surrounded by a ring and planar growth. \textbf{(c)} $\qty{2}{\micro\metre}\times\qty{2}{\micro\metre}$ AFM image of triangular etch pits on vicinal \ce{In_{0.4}Ga_{0.6}As(111)}A. The etch pits are defined by the slowest etching $\{111\}A$ faces. Reprinted figures from \citeauthor{tuktamyshevLocalDropletEtching2024}, Applied Surface Science, \textbf{669}, 160450 (2024), (CC BY) license (Ref.~\onlinecite{tuktamyshevLocalDropletEtching2024}). \textbf{(d)} AFM linescans of a nanohole etched in the same conditions as the growths in \textbf{(a)}, but with the substrate temperature at \qty{630}{\degreeCelsius}. The perpendicular linescans demonstrate the remains of a droplet recessed in one corner of the etch pit. Because the droplet density decreases with substrate temperature (see Sec.~\ref{sec:temp_dep}), the droplets formed at \qty{630}{\degreeCelsius} are relatively large. In this case the droplets were large enough that they required longer than the \qty{180}{\second} allocated to etching to be fully consumed.}\label{fig:droplet_etching}
	\end{figure*}
	
	\subsection{Equilibrium thermodynamics} \label{sec:theordynamics}
	\subsubsection{Equilibrium etching shapes}
	In principle, the equilibrium shape of the droplet etched nanoholes can be calculated from the Jaccodine-Wulff plot for the epilayer material as etched by the droplet material.~\cite{jaccodineUseModifiedFree1962} Full polar plots of the etching rates for common III-V semiconductors do not exist to our knowledge, but nonetheless Jaccodine-Wulff plots can be approximated with knowledge on the etching rate of only a few low-index facets.~\cite{xingTransientStableProfiles2017} For the common III-V semiconductors, the etching rate on the $\{100\}$ and polar $\{111\}$ facets govern the equilibrium shape of the etch pit. The etching rate is slowest on the group-III terminated $\{111\}$A facets, with higher etching rates on the more volatile group-V terminated $\{111\}$B and non-polar $\{100\}$ facets.~\cite{gatosCharacteristics111Surfaces1960} The equilibrium shape is governed by the slowest etching facets, and thus the $\{111\}$A facets should largely define the etch pits. This is strikingly demonstrated in Fig.~\ref{fig:droplet_etching}(c) by \citeauthor{tuktamyshevLocalDropletEtching2024}~\cite{tuktamyshevLocalDropletEtching2024} where the $(111)$A surface of \ce{In_{0.4}Ga_{0.6}As} samples are Ga droplet etched, producing tetrahedral etch pits bounded by well-defined $\{111\}$A facets. Far more common is droplet etching on a $\{100\}$ surface, in which case a pyramidal etch pit is expected, bounded by opposing $\{111\}$A and B facets.~\cite{vonkFacetingLocalDropletetched2018} Derivative mappings of sufficiently high resolution AFM scans can tentatively identify the facets defining the nanoholes,\cite{fusterFundamentalRoleArsenic2014, vonkFacetingLocalDropletetched2018} though tip convolution effects (especially at the nanohole bottom) must be borne in mind.  Figure \ref{fig:droplet_etching}(a) shows perpendicular linescans and their corresponding slopes for a nanohole etched at \qty{600}{\degreeCelsius}. It can be seen from the lower derivative map that the facets of the nanohole approach the equilibrium $\{111\}$ planes. In general, the facets bounding the nanoholes become more well-defined and distinguishable as the growth temperature increases up to the point of incongruent substrate evaporation.~\cite{vonkFacetingLocalDropletetched2018}
	
	As \citeauthor{vonkFacetingLocalDropletetched2018}\cite{vonkFacetingLocalDropletetched2018} have pointed out, it is ultimately the kinetics, limited by the growth temperature, that determines whether the equilibrium shape described above is reached by the system. Thus, in Fig.~\ref{fig:droplet_etching}(a) we also demonstrate the perpendicular linescans and corresponding derivative maps for a nanohole etched at \qty{550}{\degreeCelsius}. It is clear that from these graphs that the nanohole is bounded by higher-index planes. This agrees with the general observation that as the etching temperature decreases, the index of the planes defining the nanohole increases and their intersections become less well-defined, producing nanoholes with rounder openings.~\cite{heynScalingStructuralCharacteristics2014, heynDynamicsMassTransport2015} Furthermore, as the angle of inclination of the facets to substrate surface decreases, the depth of the nanoholes also decreases (as can be clearly seen in Fig.~\ref{fig:droplet_etching}(a)). The depth of the nanoholes as a function of growth temperature are well modelled by a phenomenological Arrhenius equation, with Al droplet etched nanoholes on \ce{Al_{0.35}Ga_{0.65}As} ranging from approximately \qtyrange{10}{100}{\nano\metre} in a temperature range of \qtyrange{570}{680}{\degreeCelsius}.~\cite{heynScalingStructuralCharacteristics2014} An extension of this analysis relates the nanohole depth to the aspect ratio (diameter over depth) and thus to the original size of the droplets on the surface.~\cite{tuktamyshevLocalDropletEtching2024}

	The species of the droplet material will also impact whether the equilibrium shape of the etch pits are reached by the system for any given growth. In principle, this can be rationalized from the equilibrium solubility of the III-V epilayer in a solution of the group-III material, though available data is limited to binary compounds~\cite{hallSolubilityIIICompound1963, ansaraBinaryDatabaseIII1994} and does not cover the ternary epilayers typically employed in DEE. Correspondingly few works have compared etching with different droplet species in the DEE environment,~\cite{stemmannLocalDropletEtching2008,deutschLocalDropletEtching2025} with more data required to usefully compare the etching rates of different droplet species. Rather, consideration of the droplet species is generally limited to its effect on the ring grown around the nanohole inasmuch as the ring is an alloy of the droplet and epilayer material. This has historically motivated the use of Al as an etchant, with the ring material being Al-rich and therefore optically inactive\cite{heynHighlyUniformStrainfree2009,massonEngineeringNanoholeEtchedQuantum2026} unlike with Ga- or In-droplet etching.~\cite{liangEnergyTransferUltralow2008, leeSuperLowDensity2008, stemmannLocalEtchingNanoholes2009}\\ 
	
\subsubsection{Preferential interface nucleation}\label{sec:PIN}
Preferential interface nucleation, developed by \citeauthor{wacaserPreferentialInterfaceNucleation2009},~\cite{wacaserPreferentialInterfaceNucleation2009} is a generalization of the thermodynamic theory of vapor-liquid-solid (VLS) growth of nanowires. Any VLS-like system is one in which at least three materials, or phases, of the growth material participate in the growth process. These are identified as the supply phase ($s$), a collector phase often associated with a droplet ($c$), and the final crystalline phase ($k$) of the material. Typically, the supply phase is identified with the flux of growth material to the collector from the vapor and adsorbed species, however in droplet etching this is fulfilled by the crystalline phase. Nonetheless, the supply phase remains an integral part of droplet etching, as seen below. For simplicity, we will consider the case of Ga droplet etching on a GaAs epilayer. This allows us to use the simpler expressions for homogeneous nucleation, though the qualitative results of this analysis should be readily applicable to the more common Al droplet etching on AlGaAs. We note that the same does not apply to droplet etching with In, which often exhibits complicated strain-driven effects and material intermixing.~\cite{deutschTelecomCbandPhoton2023, deutschLocalDropletEtching2025}

According to \citeauthor{wacaserPreferentialInterfaceNucleation2009},~\cite{wacaserPreferentialInterfaceNucleation2009} the dominant thermodynamic factor that determines nucleation, and thus growth, in any VLS-like system is the minimization of the Gibbs free energy of nucleation ($\Delta G$). The Gibbs free energy is in turn dependent on the supersaturation ($\Delta \mu_{ij} = \mu_i - \mu_j$ where $\mu_{i}$ is the chemical potential of GaAs in phase $i$) and interfacial energy of a nucleus created between the various phases. In the initial transient of droplet etching, the supply and crystalline phases are supersaturated with respect to the collector ($\Delta\mu_{sc}, \Delta\mu_{kc} > 0$), and there is a small amount of etching,\cite{fusterFundamentalRoleArsenic2014} as illustrated in Fig.~\ref{fig:droplet_etching}(b). However, the equilibrium solubility of GaAs in liquid Ga is low ($<1\%$ at \qty{600}{\degreeCelsius}),~\cite{rubensteinSolubilitiesGaAsMetallic1966} and so $\Delta\mu_{sc}$ and $\Delta\mu_{kc}$ quickly go to zero and etching stops without a process to remove As from the droplet. The fact that rings are observed around the nanoholes is a strong indication that the three-phase boundary (TPB) between vapor, droplet, and epilayer is a minimum in the Gibbs free energy and a sink for the As in the droplet. The Gibbs free energy at the TPB in the droplet etching system is
\begin{equation}\label{eq:gibbs_TPB}
	\Delta G_{\text{TPB}} = -n\Delta \mu_{sk} + P_{ck}t\sigma_{ck} + P_{sk}t\sigma_{sk}
\end{equation}
where $n$ is the number of atoms, $t$ is the edge height of a small nucleus formed at the boundary, $P_{ij}$ is the perimeter length of the nucleus formed between phases $i$ and $j$, and $\sigma_{ij}$ is the specific surface energy of the $i$-$j$ interface. In Eq.~(\ref{eq:gibbs_TPB}), $\Delta\mu_{sk}$ is identified as the remaining, dominant, supersaturation term driving nucleation. The TPB therefore removes As at the droplet edge and drives a concentration gradient in the droplet itself. This concentration gradient maintains a supersaturation in the crystalline phase at the droplet-epilayer interface, and sustains etching.

There are a couple of consequence that can immediately be derived from Eq.~(\ref{eq:gibbs_TPB}). Firstly, if there is no flux of As either in the vapor or adsorbed to the surface, then $\Delta\mu_{sk} \leq 0$ and nucleation at the TPB should be energetically unfavorable, with the further consequence that the droplets will saturate and etching will not occur. This is precisely what is observed for DEE growths where the As BEP $\leq \qty{E-9}{\torr}$. In such growths, droplets etch a small amount of the epilayer to achieve an equilibrium solubility of As, but are otherwise conserved on the surface.~\cite{fusterFundamentalRoleArsenic2014, heynRoleArsenicAluminum2016} The same effect is achieved by over-heating the sample in the absence of an As overpressure, depleting the surface reconstruction of As.~\cite{heynRoleArsenicAluminum2016} Secondly, if $\sigma_{ck} > \sigma_{sk}$, then the Gibbs free energy of nucleation on the crystal away from the droplet
\begin{equation}\label{eq:gibbs_sk}
	\Delta G_{sk} = -n\Delta \mu_{sk} + P_{sk}t\sigma_{sk}
\end{equation}
will be less than $\Delta G_{\text{TPB}}$. This suggests that planar growth between diffused droplet material and As from the supply phase is also important, and even dominates if the kinetics is not limited. This is observed experimentally wherein electrical current density measurements reveal an epitaxial AlAs layer after Al droplet etching,~\cite{heynDynamicsMassTransport2015} and geometric arguments based off droplet, ring, and nanohole volumes estimate upwards of 50\% of the original droplet is consumed in planar growth outside the droplets.~\cite{atkinsonIndependentWavelengthDensity2012} The basic processes governing the redistribution of material after the initial etching transient are depicted in Fig.~\ref{fig:droplet_etching}(b). Thus, the requirement that there be a reservoir of the group-V element for droplet etching to occur, and the existence of the ring around the nanoholes, is seen from Eq.~(\ref{eq:gibbs_TPB}) and Eq.~(\ref{eq:gibbs_sk}) to be a consequence of the thermodynamics of the system.
	
\subsection{Kinetics of etching}
\subsubsection{Monte Carlo based approaches} \label{sec:monte_carlo}
Monte Carlo simulations, which model the energetics of individual atomistic interactions, provide the most detailed insight into droplet etching, however, computation time limits these simulations to lower temperatures ($\lessapprox \qty{400}{\degreeCelsius}$). The currently available works are simulations that specifically model Ga droplets on the GaAs surface.~\cite{reyesUnifiedModelDroplet2013, vasilenkoFormationGaAsNanostructures2015} The key rates predicted by these simulations are the rate of substrate etching by the droplet and the rates of crystallization at the droplet-epilayer interface, the three-phase boundary, and on the epilayer away from the droplet.

In both works, the dominant pathway for the consumption of the droplet and liquified epilayer material in the DEE regime is the ``wicking''~\cite{reyesUnifiedModelDroplet2013} out of the material. Wicking is defined here as a step-flow growth mode in which it is energetically favorable for the droplet material to crystallize with surface As outside the immediate vicinity of the droplet. The result is a disk of material around a droplet/nanohole which is readily seen in AFM scans.~\cite{sanguinettiDropletEpitaxyNanostructures2018} The diffusion radius for wicked material outside the droplet is governed by the effective distance $l$ travelled by Ga atoms which escape from the droplet, and is\cite{biettiGalliumSurfaceDiffusion2014} 
\begin{equation}\label{eq:diff_len}
	l = \sqrt{D_0 \exp{(-E_A / k_B T)}\frac{N_s}{J_{\text{As}}}}
\end{equation} 
where $E_A$ is the activation energy for surface diffusion, $D_0$ the diffusivity prefactor, $N_s$ the number of surface sites, and $J_{\text{As}}$ the As flux. If $l$ is greater than the average distance between droplets, than wicking transitions to a layer-by-layer growth mode. In such a case, kinetics no longer inhibits the preferential nucleation of GaAs in planar growth, and clearly ties into Eq.~(\ref{eq:gibbs_sk}) and the theory of preferential interface nucleation above. From Eq.~(\ref{eq:diff_len}), the transition to layer-by-layer growth occurs at sufficiently high substrate temperature or low As overpressure. For Ga droplets on \ce{Al_{0.3}Ga_{0.7}As}, the transition to layer-by-layer growth of droplets occurs at lower temperatures and higher As overpressures than used in DEE.~\cite{leeEvolutionSelfassembledSingle2006} For Al droplets on GaAs, diffusion disks can persist up to \qty{500}{\degreeCelsius}, but only under an As overpressure $\geq \qty{1E-6}{\torr}$.~\cite{liHoledNanostructuresFormed2010} Under low group-V overpressures typical of DEE, diffusion disks are not seen for droplet etching on AlGaAs, however the same is not true for droplet etching on \ce{AlGaSb} where diffusion rings around nanoholes are readily observed.~\cite{hilskaNanoholeEtchingAlGaSb2021, chelluHighlyUniformGaSb2021} As in Sec.~\ref{sec:PIN}, electrical current density and nanohole morphology measurements support the existence of a thin epitaxial layer derived from the droplet material in typical DEE GaAs QD growth.~\cite{heynDynamicsMassTransport2015, atkinsonIndependentWavelengthDensity2012}

Both works also predict that the volume and height of the ring formed around the nanoholes depends sensitively on the rates of wicking and crystallization at the TPB and droplet-crystal interface. From the atomistic perspective, the ring forms because the TPB has the highest local concentration of As. The As at the TPB arrives directly from the vapor, from diffusion on the surface, and from diffusion within the droplet,~\cite{vasilenkoFormationGaAsNanostructures2015} and justifies the claim that $\Delta\mu_{sk}$ is the dominant supersaturation term in Eq.~(\ref{eq:gibbs_TPB}). The rates of crystallization are particularly sensitive to the As overpressure during droplet etching as this controls the rate of diffusing As that coordinates with droplet material. One consequence is that the ring formed around the nanohole from droplet crystallization should increase in volume and height as the As overpressure increases (however see below in Sec.~\ref{sec:etching_rate_eq}). A second consequence is that, with increased diffusion of As into the droplet, the droplet can supersaturate with As and regrowth occurs on the droplet-crystal interface, reducing the depth of the nanoholes. This is attested experimentally for Ga droplet etching on \ce{Al_{0.36}Ga_{0.64}As}, where nanohole depths reduce from \qtyrange{10}{3}{\nano\metre} when increasing the As BEP from \qtyrange{1E-7}{3E-6}{\torr}.~\cite{heynInfluenceGaCoverage2009} Regrowth on the nanoholes during etching has also been predicted by \citeauthor{vonkFacetingLocalDropletetched2018}\cite{vonkFacetingLocalDropletetched2018} for Al droplet etching of AlGaAs on the basis of the phase diagram for the Al-Ga-As system.~\cite{liThermodynamicReassessmentAlAsGa2001}

\subsubsection{Rate-equation approaches} \label{sec:etching_rate_eq}
To model the evolution of etching at the higher temperatures typically employed in DEE ($\geq \qty{500}{\degreeCelsius}$), rate equation approaches based on kinetic models of droplet etching were developed.~\cite{heynKineticModelLocal2011, liOriginNanoholeFormation2014} These models specifically target the simplest case of Ga droplet etching on GaAs. The rate equation approach of \citeauthor{heynKineticModelLocal2011}~\cite{heynKineticModelLocal2011} makes predictions on the nanohole morphology qualitatively consistent with the Monte Carlo simulations, but assumes a core-shell model geometry to the droplets throughout etching. The core-shell concept is collaborated for the lower temperature, higher As overpressure, droplet epitaxy growths,~\cite{kumahAtomicscaleMappingQuantum2009} however there is currently no evidence that it persists under DEE growth conditions. \citeauthor{liOriginNanoholeFormation2014}~\cite{liOriginNanoholeFormation2014} assume that the droplet etching is governed by a temperature activated etching rate, with recrystallization of the droplet limited by the diffusion of As to the TPB and on the surface of the epilayer, an assumption also successfully applied to modelling droplet etching on AlGaSb.~\cite{hilskaNanoholeEtchingAlGaSb2021} This assumption entails the existence of a sufficient population of Ga adatoms on the surface to saturate the As flux to the epilayer, and is justified by rate equation models in which adatom detachment from droplets is a temperature activated process.~\cite{ratschNucleationTheoryEarly2003, heynModelingGaDroplet2021} In general, it is difficult to assess the validity of the rate equation approaches because some of the key predictions on nanohole morphology that differentiate it from Monte Carlo simulations are not generally recorded by experimenters. As an example, the volume and height of the ring crystallized around the nanoholes is predicted by the rate equation approaches to decrease with an increase As overpressure. The opposite is predicted by the Monte Carlo simulations (see Sec.~\ref{sec:monte_carlo}). The fact that the ring is predicted to be an alloy of the droplet and substrate material holds consequences for the confining potential, and thus photoluminescence properties of the DEE QDs.~\cite{vonkFacetingLocalDropletetched2018} It is therefore worthwhile measuring the geometry of such features, and verifying or disputing the assumptions of the rate equation and Monte Carlo approaches.
	
\subsection{Other observations on droplet etching}
It is usual practice in DEE growths to allow \qtyrange{2}{5}{\minute} of growth interrupt during the etching step to provide sufficient time for etching and complete consumption of the droplet material.~\cite{heynHighlyUniformStrainfree2009, atkinsonIndependentWavelengthDensity2012, huoUltrasmallExcitonicFine2013, kruckCriticalAluminumEtch2024} The substrate temperature during this time is typically kept the same as that in droplet deposition, but the group-V overpressure can often be increased during this step, sometimes in a time-dependent manner.~\cite{dasilvaGaAsQuantumDots2021} Because an increase in the group-V overpressure during etching increases the rate of consumption of droplet material, for sufficiently high overpressure, etching completes within a few seconds and the effect of the extended growth interrupt is to anneal the nanoholes, gradually reducing their depth.~\cite{fusterFundamentalRoleArsenic2014}
	
Extended growth interrupts are strictly only necessary under conditions when etching, and the lifetime of the droplet on the surface, are prolonged. Such cases arise when either the group-V overpressure is very low, or when etching is concurrent with droplet nucleation and coarsening under conditions of either low droplet deposition flux or high droplet material coverage.~\cite{heynDynamicsSelfassembledDroplet2009} Under all such conditions of prolonged etching, the asymmetry in the droplet and its ring becomes exaggerated, and the ring often develops a distinctive tetrahedral growth directed along $[\bar{1}10]$. The time resolved experiments of \citeauthor{fusterFundamentalRoleArsenic2014}~\cite{fusterFundamentalRoleArsenic2014} demonstrate that under conditions of prolonged etching, the droplet will preferentially etch into one corner of the nanohole as defined by its facets. This type of etching has its origin in the droplet shape cycles observed by \citeauthor{shorlinShapeCycleGa2007}~\cite{shorlinShapeCycleGa2007} for Ga on GaAs, wherein the Ga droplets cyclically etch rectangular pits and then retract into a corner of the pit to reduce their interfacial energy with the substrate. This behavior, in combination with the preferential growth on B-facets~\cite{atkinsonIndependentWavelengthDensity2012, fusterFundamentalRoleArsenic2014} and the anisotropic diffusion in $[\bar{1}10]$ for Ga/Al adatoms on the GaAs$(2\times 4)$ surface reconstruction~\cite{kleyNovelDiffusionMechanism1997, biettiGalliumSurfaceDiffusion2014, wangMechanismAluminumDroplet2023} produces the distinctive anisotropy in the ring, and can contribute to the asymmetry of the nanohole itself. Hence, the nanoholes become narrower along $[\bar{1}10]$ because the wicked out droplet material has preferentially diffused and recrystallized on the B-facets. Figure \ref{fig:droplet_etching}(d) demonstrates a nanohole in which etching had not completed, and the droplet had retracted into a corner of the etch pit; see also the figures in \citeauthor{fusterFundamentalRoleArsenic2014}.~\cite{fusterFundamentalRoleArsenic2014}
	
\section{Regrowth}\label{sec:redeposition}
The final step of DEE is the regrowth of a lower bandgap material into the etched nanohole templates, and their subsequent capping with higher bandgap material creating QDs. There exist two modes of regrowth into the nanohole templates: strain-free growth as exemplified by GaAs quantum dots in \ce{AlGaAs} nanoholes, and strained (or SK) growth realized by InAs quantum dots in GaAs nanoholes. 
	
	\subsection{Strain-free regrowth}\label{sec:strain-free_regrowth}
	\begin{figure}
		\centering
		\includegraphics[width=\columnwidth]{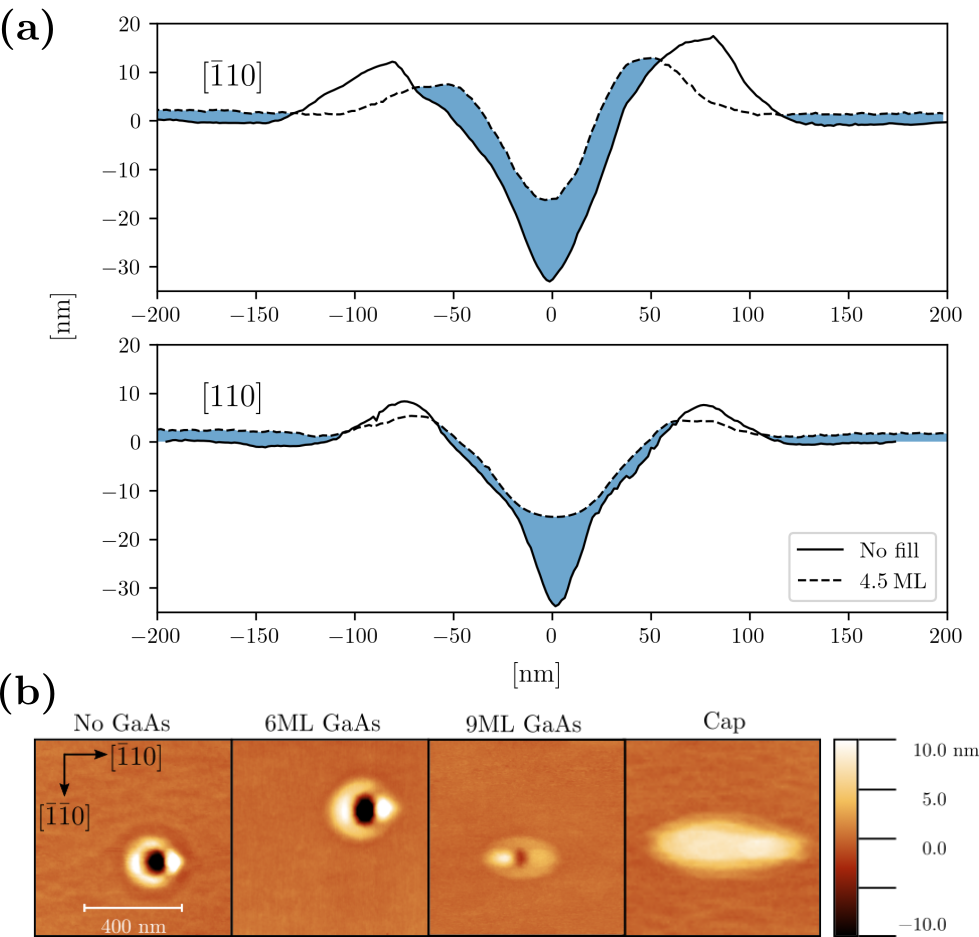}
		\caption{\textbf{(a)} Representative AFM linescans in $[\bar{1}10]$ and $[110]$ of a single nanohole with \qty{4.5}{\monolayer} of GaAs regrowth. The linescans come from two separate growths; these growths were the same in all respects except that the growth represented by the dashed linescans had a final regrowth of \qty{4.5}{\monolayer} of GaAs on the surface. The growth conditions were: substrate temperature \qty{600}{\degreeCelsius}, group-V overpressure \qty{6E-8}{\torr}, droplet deposition flux \qty{0.24}{\monolayer\per\second}, and GaAs growth rate \qty{0.25}{\monolayer\per\second}. The relative surface levels of the two scans has been set assuming a difference of \qty{4.5}{\monolayer} GaAs far from the nanoholes. \textbf{(b)} Representative AFM scans of GaAs regrowth into Al-etched nanoholes in \ce{Al_{0.33}Ga_{0.67}As}. GaAs was regrown into the nanoholes at a temperature of \qty{600}{\degreeCelsius} and a growth rate of \qty{0.25}{\monolayer\per\second} with an \ce{As_2} BEP of \qty{2.6E-6}{\torr}. Regrowth is seen to preferentially occur on the B-type step edges of the nanohole facet. This regrowth behavior continues with the \ce{Al_{0.33}Ga_{0.67}As} capping layer (which is \qty{80}{\nano\metre} thick in the rightmost scan), producing asymmetric hillocks on the surface with allow for easy identification of the sites of buried QDs.}\label{fig:hole_regrowth}
	\end{figure}
	
	Regrowth in strain-free systems has only been studied in depth for GaAs in \ce{AlGaAs} nanoholes, though it is probably the case that the conclusions derived for this system are applicable to regrowth in other strain-free or nearly strain-free III-V systems. The simplest model for the regrowth of GaAs into the nanohole templates may be termed as `bottom-up', wherein the Ga adatoms diffuse until they reach the bottom of a nanohole. Here the Ga becomes locked into the crystal lattice, and the QD shape is that of the nanohole with a truncation in the growth direction determined by the total regrowth amount. This model of regrowth has been assumed by numerous groups in their growth of GaAs QDs,~\cite{heynHighlyUniformStrainfree2009, huoUltrasmallExcitonicFine2013, kusterDropletEtchingDeep2016, huberHighlyIndistinguishableStrongly2017,zhaiQuantumInterferenceIdentical2022} with a growth interrupt universally applied after deposition of GaAs regrowth in order to promote Ga diffusion into the nanoholes. However, as the AFM linescans in Fig.~\ref{fig:hole_regrowth}(a) demonstrate, the actual regrowth topology is more complicated than the bottom-up assumption. In reality the regrowth topology is a combination of growth on the facets of the nanoholes,~\cite{atkinsonIndependentWavelengthDensity2012} and capillarity-induced diffusion of adatoms to the nanohole bottoms \cite{sablonStructuralEvolutionFormation2008, heynFieldControlledQuantumDot2019} where the surface chemical potential is known to be a local minimum.~\cite{wuFabricationUltralowdensityQuantum2017} This topology was first termed by \citeauthor{heynFieldControlledQuantumDot2019}~\cite{heynFieldControlledQuantumDot2019} as ``cone shell'', and holds important consequences for the symmetry of the QDs especially when the nanohole is not completely filled in. For bottom-up growth, the geometry of a GaAs QD is the same as its nanohole template, with the QD symmetry independent of the specifics of the regrowth step. However, for cone shell QD growth, the symmetry is also determined by the growth rates on the nanohole facets. This is particularly important for regrowth on \ce{AlGaAs} since the growth rate of GaAs is different on the $\{111\}$A and B facets which approximately define the nanohole template.~\cite{atkinsonIndependentWavelengthDensity2012} Here, Ga preferentially chemisorbs onto the As-terminated B-faces, resulting in QDs which are thicker along $[\bar{1}10]$, a process clearly demonstrated in the AFM scans of Figs.~\ref{fig:hole_regrowth}(a)-(b). Thus, any regrowth short of complete infilling of the nanoholes will lead to asymmetric QDs. This asymmetry can be mitigated to some extent by choice of the molecular species for the group-V element used during regrowth. As demonstrated by \citeauthor{deutschTelecomCbandPhoton2023},~\cite{deutschTelecomCbandPhoton2023} sidewall growth is significantly enhanced when using \ce{As_2} as compared to \ce{As_4} because the diffusion length of the group-III adatoms is reduced in the presence of the dimer species.

	The cross-sections of regrown nanoholes can be also be probed with transmission electron microscopy (TEM), though the chances of successfully locating a QD with this method are small due to the very low densities realized in DEE. TEM of single QDs confirms the strain-free nature of DEE and demonstrates negligible material intermixing between nanohole and regrowth.~\cite{sablonStructuralEvolutionFormation2008, nemcsicsCrosssectionalTransmissionElectron2011,leguayUnveilingElectronicStructure2024,massonEngineeringNanoholeEtchedQuantum2026} Finally, we note that the growth asymmetry extends to the capping layer, resulting in elongated hillocks which can be conveniently used for identifying the sites of buried QDs; see Fig.~\ref{fig:hole_regrowth}(b). There is no evidence either in the literature or in our own growths that these hillocks act as preferential nucleation centers when repeating droplet etching on the surface.  
	
	\subsection{Strained regrowth}\label{sec:strained}
	A novel adaption of the DEE technique is the use of Ga- or In-etched nanoholes in GaAs for the growth of strained InAs QDs. By employing sufficiently low InAs growth rates and a total coverage less than the critical amount (\qtyrange{1.5}{1.7}{\monolayer}) necessary for planar SK growth, InAs QDs can be grown exclusively in the nanoholes.~\cite{alonso-gonzalezLowDensityInAs2008, wuFabricationUltralowdensityQuantum2017} The deposited In adatoms diffuse from regions of high to low chemical potential, which in turn can be modelled as\cite{liSelectiveFormationMechanisms2013}
	\begin{equation}
		\mu = \mu_0 + \Omega\gamma_s k + \Omega E_\text{str} + \Omega \frac{d y_s(h_s)}{d h_s},
	\end{equation}
	where $\mu_0$ is the thermodynamic driving force of the crystallization process, $\Omega$ is the atomic volume, $\gamma_s$ is the surface energy density, $k$ is the local surface curvature, $h_s$ the height of the surface, and $E_{\text{str}}$ the mismatch strain energy density stored in the epitaxial material. AFM-generated surface profiles can be used to extract the surface curvature, whereupon the nanohole bottoms are seen to be local minimums in $\mu$.~\cite{wuFabricationUltralowdensityQuantum2017}  Like their unstrained counterparts, the InAs QDs also preferentially nucleate on the B-type facets of the nanoholes, and under certain growth conditions QD pairs can form on the opposing facets. The transition between single and QD pairs in the nanoholes templates can be controlled by both the total regrowth amount,~\cite{yuHighlyUniformSymmetric2019} or the extent of Ga/In intermixing on the facets via the As BEP.~\cite{alonso-gonzalezFormationLateralLow2009} The current state-of-the-art for InAs QDs in DEE nanoholes is found in the work of \citeauthor{spitzerTelecomOBandQuantum2024}~\cite{spitzerTelecomOBandQuantum2024} Here the authors make use of small Ga-etched nanoholes with depths of $\approxeq \qty{5}{\nano\metre}$ as templates for InAs growth. The small volume of these nanoholes allows them to be completely regrown with less than \qty{1.5}{\monolayer} InAs, avoiding SK QD growth on the planar regions of the sample. A \qty{6}{\nano\metre} strain-reducing capping layer of \ce{In_{0.3}Ga_{0.7}As} followed by GaAs red-shifts the emission of these InAs QDs into the telecom O-band.
	
	\section{Basic QD Optics}\label{sec:optics}
	DEE GaAs QDs are significantly larger than their SK InAs QD predecessors, often demonstrating lateral extensions in excess of the GaAs free-exciton Bohr radius. As first pointed out by \citeauthor{huberSingleparticlepictureBreakdownLaterally2019},\cite{huberSingleparticlepictureBreakdownLaterally2019} the excitonic states in DEE GaAs QDs are weakly confined in the lateral direction, with the Coulomb interaction dominating quantum confinement effects. Consequently, GaAs QDs have greater oscillator strength than InAs QDs, resulting in ground-state exciton lifetimes of $\approx\qty{250}{\pico\second}$~\cite{keilSolidstateEnsembleHighly2017, huberHighlyIndistinguishableStrongly2017, heynDotSizeDependentExcitons2022} and large lateral extensions of the wave function. The latter suggests that the excitonic wave function is less sensitive to dephasing and anisotropy within the QD, reducing the FSS of the ground-state exciton.~\cite{huberSemiconductorQuantumDots2018} 
	
	In this section we cover the basic emission properties of DEE GaAs QDs as they relate to the growth conditions. Excellent reviews of GaAs QDs as quantum light sources already exist,~\cite{gurioliDropletEpitaxySemiconductor2019, dasilvaGaAsQuantumDots2021} and we refer the interested reader to these for more information on the quantum optics. Discussion at the end of this section is reserved for DEE QDs realized in other material systems. Here we briefly describe our experimental setup for micro-photoluminescence measurements. In initial experiments, QDs were pumped with above-band excitation using a \qty{632}{\nano\metre} pigtailed laser diode. The sample was mounted inside a closed-cycle, helium-cooled cryostat (Attodry 800) maintained at \qty{4}{\kelvin} on an XY piezo stage. The cryostat is equipped with an apochromatic microscope objective ($\text{NA} = 0.81$), optimized for \qty{800}{\nano\metre}, which is used for both excitation and photoluminescence collection via a 10/90 beamsplitter. The objective focuses the beam to an approximately \qty{1}{\micro\metre} spot and is mounted on a vertical (Z-axis) piezo stage, enabling selective excitation of individual QDs. The excitation laser is linearly polarized, however no polarization selection was employed for above-band pumping. The emitted PL is coupled into a single-mode fiber and refocused into a \qty{750}{\milli\metre} focal-length spectrometer (HR750). Spectral dispersion is achieved using a $1200\:\text{lines/mm}$ grating, and detection is performed with a wavelength-calibrated silicon CCD, yielding a spectral resolution of \qty{40}{\pico\meter}.
	
	\subsection{Macro-photoluminescence}
	Macro-photoluminescence (macro-PL) experiments involve the excitation of a QD ensemble ($\sim \numrange{E6}{E10}$ QDs) and are a non-destructive method of assessing the average QD emission energy and sample homogeneity.~\cite{babinFullWaferProperty2022} For a DEE GaAs QD sample, two features tend to dominate the emission spectrum. The first is the emission from the GaAs quantum well co-deposited during the regrowth step, where the emission varies between \qtyrange{680}{720}{\nano\metre} depending on the regrowth amount. The second is an inhomogeneously broadened peak from the QD ensemble, with peak emission correspondingly ranging from \qtyrange{710}{800}{\nano\metre}. QD-like features in the \qtyrange{710}{720}{\nano\metre} range can be due to quantum well roughened QDs or DEE QDs. In the literature, additional QD peaks appear with high pump power at lower wavelengths and are attributed to emission from higher energy charged excitonic states.~\cite{heynHighlyUniformStrainfree2009, heynFieldControlledQuantumDot2019, babinFullWaferProperty2022, kruckCriticalAluminumEtch2024}
	
	\begin{figure*}
		\centering
		\includegraphics[width=\textwidth]{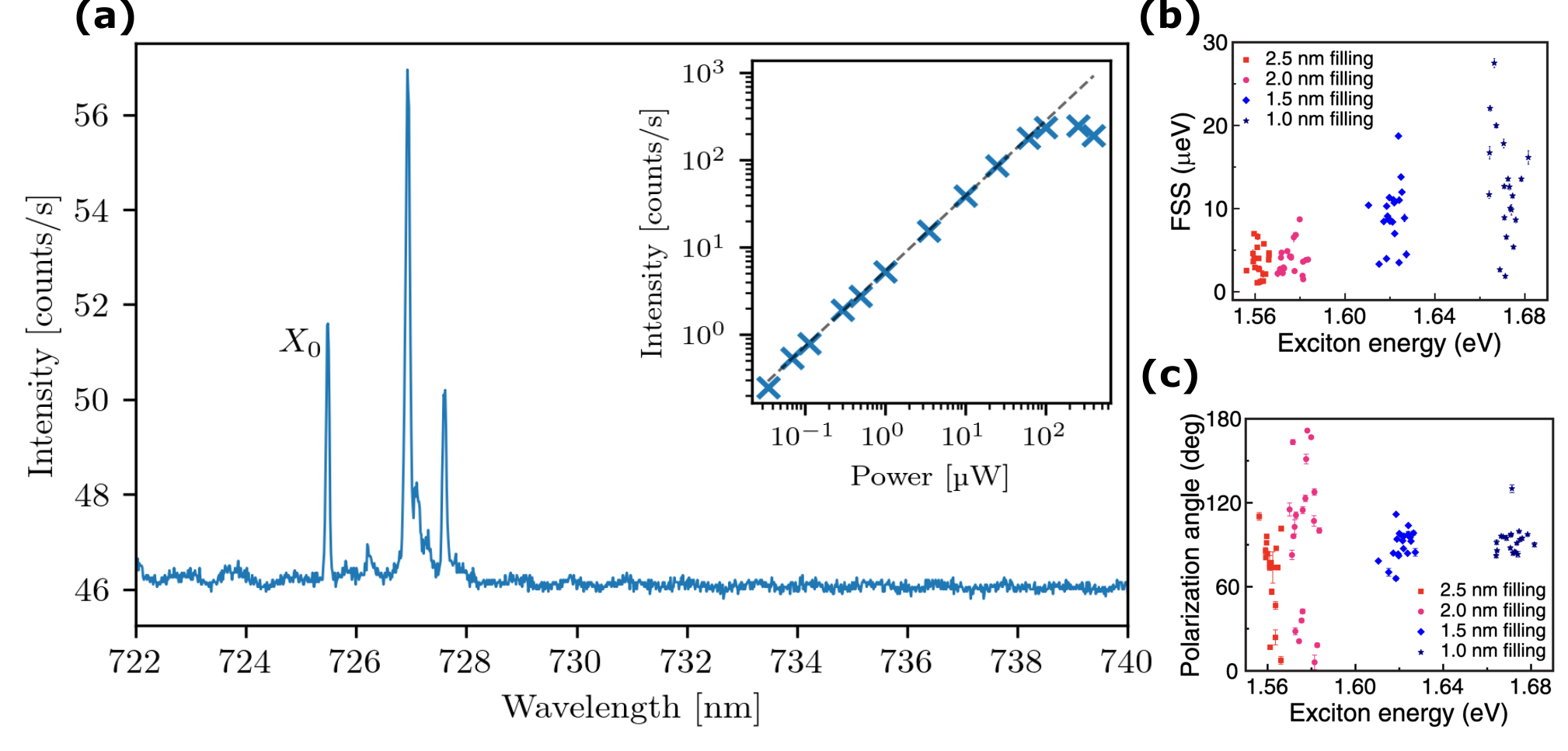}
		\caption{\textbf{(a)} \qty{4}{\kelvin} micro-photoluminescence experiment on a single GaAs QD using \qty{1}{\micro\watt} of \qty{632}{\nano\metre} above-band pumping. The growth parameters for this QD sample were: Al deposition flux $F = \qty{1.2}{\monolayer\per\second}$, substrate temperature $T = \qty{600}{\degreeCelsius}$, As overpressure $F_{\text{As}} < \qty{8E-9}{\torr}$, and a GaAs regrowth amount of \qty{1.5}{\monolayer}. The spectrum is typical of a single GaAs DEE QD. The neutral exciton ($X_0$) is identified by unity slope in a pump power dependence experiment; this is demonstrated in the inset of the figure, with saturation of the line occurring at pump powers $> \qty{100}{\micro\watt}$. Other dominant features in the spectrum are identified with ground and excited state trions.~\cite{huberSingleparticlepictureBreakdownLaterally2019} \textbf{(b)} - \textbf{(c)} FSS and polarization angle of QD emission as a function of exciton energy and nominal GaAs regrowth amount. The polarization angles in \textbf{(c)} represent the polarization direction of the high energy component of the bright exciton doublet with respect to the $[110]$ crystal direction. Reprinted figures with permission from \citeauthor{huoVolumeDependenceExcitonic2014}, \href{https://doi.org/10.1103/PhysRevB.90.041304}{Phys. Rev. B, \textbf{90}, 041304(R), 2014}. Copyright 2014 by the American Physical Society (Ref.~\onlinecite{huoVolumeDependenceExcitonic2014}).}\label{fig:micro-PL}
	\end{figure*}
	
	\subsection{Micro-photoluminescence}
	The spectrum of a single GaAs QD from a typical \qty{4}{\kelvin} micro-photoluminescence experiment with above-band pumping is given in Fig.~\ref{fig:micro-PL}a. The neutral exciton ($X_0$) is identified through its linear pump-power dependence (on a log-log scale) along with its linear polarization.~\cite{santoriTriggeredSinglePhotons2001} It is separate from a cluster of longer wavelength peaks that include a charged exciton peak, and excited-state related peaks.~\cite{huberSingleparticlepictureBreakdownLaterally2019} While the biexciton state is assumed to be present, it is not generally observed (see e.g. Ref.~\onlinecite{huberSingleparticlepictureBreakdownLaterally2019}) with exception from \citeauthor{heynSingledotSpectroscopyGaAs2010}~\cite{heynSingledotSpectroscopyGaAs2010,  heynDotSizeDependentExcitons2022} and \citeauthor{grafExcitonicStatesGaAs2014}~\cite{grafExcitonicStatesGaAs2014} where it is observed $ \approx \qty{2}{\nano\metre}$ longer than the exciton. In our lab, we have not observed emission from the biexciton in GaAs QDs with above-band pumping.
	
	For DEE QDs, the emission energy of the neutral exciton ($E_{X_0}$) is dependent on the amount of regrowth into the nanoholes, with the average emission energy following an empirical quadratic dependence,~\cite{atkinsonIndependentWavelengthDensity2012, grafExcitonicStatesGaAs2014, heynFieldControlledQuantumDot2019, huoVolumeDependenceExcitonic2014} consistent with quantum confinement effects. For completely filled QDs with a constant amount of nominal GaAs regrowth, \citeauthor{loblCorrelationsOpticalProperties2019}\cite{loblCorrelationsOpticalProperties2019} report  $E_{X_0} \sim 1/d^{0.556}$, where $d$ is the nanohole depth. We note that this result relies on a phenomenological approach for relating the experimentally accessible Voronoi cell area of the QD to the nanohole depth.
	
	The FSS of the ground-state exciton is well-known to be dependent on the symmetry of the QD.~\cite{bayerFineStructureNeutral2002, huberSemiconductorQuantumDots2018} For DEE QDs, the symmetry of the quantum dot is dependent on the in-plane symmetry of the nanohole, and the anisotropic growth rate on the nanohole facets (see Sec.~\ref{sec:redeposition}). For example, \citeauthor{huoUltrasmallExcitonicFine2013}~\cite{huoUltrasmallExcitonicFine2013} have demonstrated a significant reduction in the FSS for nanohole templates exhibiting circular in-plane symmetry; from a FSS of \qty{47}{\micro\electronvolt} for deep, partially regrown, anisotropic nanoholes, to \qty{3.5}{\micro\electronvolt} for shallower, largely regrown, symmetric nanoholes. The dependence of the FSS on the growth conditions can also be inferred from the distribution of the single-particle wave functions of the exciton within the QD. Simulations based off eight-band $\mathbf{k}\cdot\mathbf{p}$ theory predict the single-particle wave functions of the electron and hole reside at the center of mass of the DEE QDs.~\cite{huoVolumeDependenceExcitonic2014, heynFieldControlledQuantumDot2019} These wave functions strongly overlap and hence the exciton has a large transition dipole moment. Furthermore, the aspect ratio of the wave functions decreases as the height increases in geometrically similar QDs.~\cite{huoVolumeDependenceExcitonic2014} Thus, the excitonic wave function is more sensitive to lateral anisotropies of the QD as the regrowth amount decreases, a phenomenon which is only exacerbated by the anisotropic growth rates on the nanohole facets demonstrated in Sec.~\ref{sec:redeposition}. This has been concisely shown by \citeauthor{huoVolumeDependenceExcitonic2014}~\cite{huoVolumeDependenceExcitonic2014} wherein there is a reduction in the average FSS from \qty{13.8}{\micro\electronvolt} for partially regrown nanoholes realized by \qty{1.0}{\nano\metre} regrowth, to \qty{1.6}{\micro\electronvolt} for fully regrown nanoholes realized by \qty{2.5}{\nano\metre} regrowth; see Fig.~\ref{fig:micro-PL}(b). The authors further postulate that the FSS is affected by weakly randomizing effects such as the irregular shape of the ring around the nanoholes and alloy fluctuations in the barrier.
	
	Other properties of the DEE QD emission are also directly impacted by the symmetry of the nanohole and volume of regrowth. \citeauthor{heynDotSizeDependentExcitons2022}~\cite{heynDotSizeDependentExcitons2022} have demonstrated an increase in lifetime of the ground-state exciton from \qty{370}{\pico\second} to \qty{580}{\pico\second} for cone-shell QDs with nominal regrowths of $\qty{1}{\monolayer}$ to \qty{2}{\monolayer} GaAs, respectively. Similarly, the ground-state dark exciton increases in lifetime from \qty{3.2}{\nano\second} to \qty{6.7}{\nano\second}.~\cite{heynDotSizeDependentExcitons2022} The result for the bright exciton is contrary to what would be expected as the size of the quantum emitter increases, however the increase in lifetime of the dark exciton is probably indicative of a decrease in non-radiative recombination centers from crystal defects.~\cite{heynDotSizeDependentExcitons2022} Anisotropic QDs also demonstrate highly polarized emission in the plane of growth. As shown in Fig.~\ref{fig:micro-PL}(c), the higher energy component of the bright exciton doublet is polarized in $[\bar{1}10]$ for anisotropic or underfilled nanoholes, consistent with the final QD shape anisotropy, whereas the polarization distribution is random for symmetric QDs.~\cite{huoUltrasmallExcitonicFine2013, huoVolumeDependenceExcitonic2014} There is tentative evidence (from spectrometer resolution-limited experiment data) to suggest that the full width at half maximum (FWHM) of the exciton line also decreases as the QD approaches ideal symmetry,\cite{huoUltrasmallExcitonicFine2013} though the excellent optical quality of DEE QDs generally places the FWHM below the resolution limit of most micro-PL configurations.

	Though complete infilling of the nanoholes generally leads to QDs with advantageous optical qualities, there are detriments to excessive overfilling of the nanoholes. Recent work by \citeauthor{massonEngineeringNanoholeEtchedQuantum2026}~\cite{massonEngineeringNanoholeEtchedQuantum2026} on GaSb QDs has demonstrated that overfilled nanoholes are strongly coupled to the QW co-deposited during regrowth, producing hybrid QD-QW nanostructures. Carriers produced in the QW are efficiently transferred into the QDs on short timescales ($<\qty{100}{\pico\second}$) resulting in poorly defined excitonic resonances and featureless spectrums. In contrast, deeper nanoholes with less regrowth demonstrate spectrometer resolution-limited excitonic features.~\cite{massonEngineeringNanoholeEtchedQuantum2026} Early work on DEE GaAs QDs similarly demonstrated broad, featureless, QD spectrums from overfilled shallow nanoholes.~\cite{heynOpticalPropertiesGaAs2009, heynSingledotSpectroscopyGaAs2010}
	
	If diminishingly small FSS is important, it is clear that DEE GaAs QDs require the growth of highly symmetric nanoholes with moderate depths that allow for complete regrowth. We end this section with a suggestion on the growth conditions necessary to produce such nanoholes. From Sec.~\ref{sec:etch}, ideal conical nanoholes are etched under conditions away from thermal equilibrium and in which droplet consumption is quick. Thus, the growth conditions required to produce conical nanoholes include lower growth temperatures ($\lessapprox \qty{600}{\degreeCelsius}$), moderate As overpressures that result in quick droplet consumption, and high droplet deposition fluxes ($\geq \qty{0.5}{\monolayer\per\second}$).~\cite{huoUltrasmallExcitonicFine2013} The optimal As overpressure throughout droplet deposition and etching is system specific, though other research groups have had success using \ce{As_4} BEPs in the range of \qtyrange{3E-7}{5E-7}{\torr} with droplet deposition fluxes > \qty{0.35}{\monolayer\per\second}.~\cite{dasilvaGaAsQuantumDots2021, babinFullWaferProperty2022} It is unclear what the optimal length of time is for allowing the droplets to etch before regrowth of the nanoholes. High quality DEE QDs are typically grown with \qty{3}{\minute} allocated to etching,\cite{heynHighlyUniformStrainfree2009, atkinsonIndependentWavelengthDensity2012, huoUltrasmallExcitonicFine2013, kruckCriticalAluminumEtch2024} though we note that extended periods of etching allow for the accumulation of impurities on the surface, and may introduce defect states into the QDs. Finally, a sufficient amount of regrowth should be used to completely fill the nanoholes, though care must be taken not to excessively overfill the nanoholes as this may lead to broad QD spectrums.

	\subsection{DEE QDs in other material systems}
	To conclude this review, we take a brief look at research growing DEE QDs in material systems other than AlGaAs/GaAs. One strong motivation for studying DEE QDs in other material systems is the extension of QD emission into the telecom O- and C-bands, where transmission dispersion and attenuation in silica fiber are minimized. AlGaSb/GaSb was one of the first systems investigated for this purpose, having developed from earlier work on Ga droplet etching of \ce{Al_{0.3}Ga_{0.7}Sb}.~\cite{hilskaNanoholeEtchingAlGaSb2021} O-band emission from GaSb QDs in Al-etched \ce{Al_{0.3}Ga_{0.7}Sb} nanoholes was first demonstrated by \citeauthor{chelluHighlyUniformGaSb2021},~\cite{chelluHighlyUniformGaSb2021} with single-QD emission later confirmed by \citeauthor{michlStrainFreeGaSbQuantum2023}~\cite{michlStrainFreeGaSbQuantum2023} Unlike GaAs, GaSb has a direct-to-indirect bandgap crossover at nanometer scales,~\cite{leguayUnveilingElectronicStructure2024} effectively setting a minimum size to optically active GaSb QDs. For the shallower nanoholes that were preferentially grown by \citeauthor{chelluHighlyUniformGaSb2021},~\cite{chelluHighlyUniformGaSb2021} this necessitated a large amount of regrowth, overfilling the nanoholes and likely producing QDs that were completely contiguous with the co-deposited QW. In recognition of the non-zero lattice mismatch between GaSb and AlGaSb, \citeauthor{massonEngineeringNanoholeEtchedQuantum2026}~\cite{massonEngineeringNanoholeEtchedQuantum2026} recently extended Al droplet etching to quaternary \ce{Al_{0.3}Ga_{0.7}SbAs} lattice-matched to GaSb. Deeper nanoholes were shown to be advantageous for reducing the coupling of the QDs to the GaSb QW co-deposited during regrowth, with QDs grown in deeper nanoholes demonstrating superior emission properties. The potential for these QDs to act as telecom-wavelength quantum light sources is highly promising, with an average FSS of \qty{11 \pm 5}{\micro\electronvolt} and quasi-resonant pumping of a single QD demonstrating a single-photon purity of $g^{(2)}(0) = 0.029 \pm 0.011$.~\cite{massonEngineeringNanoholeEtchedQuantum2026}
	
	\citeauthor{deutschTelecomCbandPhoton2023}\cite{deutschTelecomCbandPhoton2023} have demonstrated single-QD C-band emission for \ce{In_{0.53}Ga_{0.47}As} QDs lattice-matched to the InP substrate in \ce{In_{0.52}Al_{0.48}}-etched \ce{In_{0.52}Al_{0.48}As} nanoholes. The alloyed droplet is chosen to ensure homogeneity of the nanohole ring with the etched epilayer. The rapid etching of In droplet material routinely produces faceted nanoholes, and makes it difficult to produce symmetric nanoholes; \citeauthor{deutschTelecomCbandPhoton2023}\cite{deutschTelecomCbandPhoton2023, deutschLocalDropletEtching2025} have found it necessary to regrow up to \qty{50}{\nano\metre} of \ce{In_{0.52}Al_{0.48}As} into the nanoholes in order to retain and produce adequately symmetric nanoholes. Work on these QDs is still in the initial stages, with single-QD emission spectrums qualitatively similar to other DEE QDs (e.g. Fig.~\ref{fig:micro-PL}(a)), though quantum light emission remains to be confirmed. 
	
	\citeauthor{spitzerTelecomOBandQuantum2024}~\cite{spitzerTelecomOBandQuantum2024} realized O-band emission using strained InAs QDs, capped with an InGaAs strain-reducing layer, in Ga-etched GaAs nanoholes. Similar work has recently been reported by \citeauthor{dasilvaLowdensityInGaAsAlGaAs2025}~\cite{dasilvaLowdensityInGaAsAlGaAs2025} for \ce{In_{x}Ga_{1-x}As} regrown in \ce{Al_{0.33}Ga_{0.67}As} nanoholes, with changes in the In content allowing for adjustments of the QD emission energy in the range of $\approx \qtyrange{800}{900}{\nano\metre}$. Predecessors to these works have demonstrated single InAs QD emission in Ga-etched nanoholes in the range of \qtyrange{890}{1130}{\nano\metre} by varying the total strained regrowth from \qtyrange{0.9}{1.7}{\monolayer}.~\cite{alonso-gonzalezLowDensityInAs2008, wuFabricationUltralowdensityQuantum2017} The use of Ga as a droplet material is advantageous in obtaining symmetric nanoholes and homogeneity between the epilayer and the rings formed around the nanoholes, in turn avoiding additional optically active nanostructures.\cite{liangEnergyTransferUltralow2008,leeSuperLowDensity2008,stemmannLocalEtchingNanoholes2009} In contrast, the use of In as droplets may be useful in reducing intermixing effects between QD and nanohole, however it produces optically active quantum ring structures.~\cite{liangEnergyTransferUltralow2008, yuHighlyUniformSymmetric2019} The study of InAs QD pair formation in the nanoholes remains an interesting avenue of study as these structures are highly reproducible and have some early, ensemble signatures of lateral coupling.~\cite{alonso-gonzalezFormationLateralLow2009} \citeauthor{kusterDropletEtchingDeep2016}~\cite{kusterDropletEtchingDeep2016} have vertically stacked GaAs QDs within deep AlGaAs nanoholes, and we note that similar to SK QDs,\cite{solomonVerticallyAlignedElectronically1996, stinaffOpticalSignaturesCoupled2006} the growth of coupled QDs by DEE promises to be a rich, if currently undeveloped, area of future study.
	
	\section{Conclusion}
	DEE QD growth has clear advantages over SK QD growth, in particular, a wider single excitonic emission wavelength range, lower FSS, longer electron-spin coherence, and in general better QD growth control. We have described the growth techniques and characterization used for DEE QDs in III-V MBE, and reviewed the parameter space available to varying the topology and photoluminescence properties. All aspects of the DEE growth process can be further refined, with the research likely to be relevant to fundamental knowledge on crystal growth. For example, further research could be undertaken in understanding the origin of bimodal droplet distributions, the applicability of the Pimpinelli-Einstein capture zone approach to droplet nucleation, and the effect of the group-V overpressure on nanohole morphology. Research can also be undertaken in how the fundamental optics of these QDs relates to the growth conditions, with the results of such studies likely to inform more sophisticated quantum optics experiments. For example, is the appearance of the biexcition under above-band pumping a product of the growth technique?
	
	A clear advantage of DEE growth over SK QD growth is its applicability to other material systems. Though GaAs QDs are by far the most studied, they are limited to emission approximately in the range of \qtyrange{690}{800}{\nano\metre}. The application of DEE to other material systems, such as AlGaSb/GaSb\cite{chelluHighlyUniformGaSb2021, michlStrainFreeGaSbQuantum2023} and InAlAs/InGaAs,\cite{deutschTelecomCbandPhoton2023} has allowed quantum light emission from quantum dots in the telecom C- and O-bands. The application to more material systems yet is tantalizing; for example, QDs emitting at \qtyrange{1300}{2000}{\nano\metre} may be realized in the phosphide system. Nor is DEE limited to MBE, with recent work demonstrating droplet etching by metal-organic vapor phase epitaxy.~\cite{salaInAsInPQuantum2020, salaLocalDropletEtching2024} However, each material and epitaxy system is bound to have its own pathologies, and much work will have to be done in determining the parameter space for the optimized growth of DEE QDs.

	\stoptoc
	
\begin{acknowledgments}
The authors would like to acknowledge Nicole J. Blaess from the Defence Science and Technology Group for valuable help with the optics. We acknowledge support through the Australian Research Council (ARC) grant DP220103624, and the ARC Training Centre in Current and Emergent Quantum Technologies (CE-QuTech). 
\end{acknowledgments}
	
\section*{Author Declarations}
\subsection*{Conflict of Interest}
The authors have no conflicts to disclose.
\subsection*{Author Contributions}
\textbf{Declan Gossink}: Writing - original draft (lead); Data curation (lead); Methodology (lead); Investigation (lead). \textbf{Undurti S. Sainadh}: Writing - original draft (supporting); Data curation (supporting); Investigation (supporting); Supervision (supporting). \textbf{Glenn S. Solomon}: Conceptualization (lead); Funding acquisition (lead); Supervision (lead); Writing - review \& editing (lead). 
\section*{Data Availability}
The data that support the findings of this study are available from the corresponding author upon reasonable request.
	
\renewcommand{\bibsection}{\section*{References}}
\bibliography{MyLibrary.bib}

@article{aharonovichSolidstateSinglephotonEmitters2016,
  title = {Solid-State Single-Photon Emitters},
  author = {Aharonovich, Igor and Englund, Dirk and Toth, Milos},
  year = 2016,
  month = oct,
  journal = {Nature Photonics},
  volume = {10},
  number = {10},
  pages = {631--641},
  publisher = {Nature Publishing Group},
  issn = {1749-4893},
  doi = {10.1038/nphoton.2016.186},
  urldate = {2025-01-08},
  copyright = {2016 Springer Nature Limited},
  langid = {english}
}

@article{akopianEntangledPhotonPairs2006,
  title = {Entangled {{Photon Pairs}} from {{Semiconductor Quantum Dots}}},
  author = {Akopian, N. and Lindner, N. H. and Poem, E. and Berlatzky, Y. and Avron, J. and Gershoni, D. and Gerardot, B. D. and Petroff, P. M.},
  year = 2006,
  month = apr,
  journal = {Physical Review Letters},
  volume = {96},
  number = {13},
  pages = {130501},
  publisher = {American Physical Society},
  doi = {10.1103/PhysRevLett.96.130501},
  urldate = {2025-09-08}
}

@article{alonso-gonzalezFormationLateralLow2009,
  title = {Formation of {{Lateral Low Density In}}({{Ga}}){{As Quantum Dot Pairs}} in {{GaAs Nanoholes}}},
  author = {{Alonso-Gonz{\'a}lez}, P. and {Mart{\'i}n-S{\'a}nchez}, J. and Gonz{\'a}lez, Y. and Al{\'e}n, B. and Fuster, D. and Gonz{\'a}lez, L.},
  year = 2009,
  month = may,
  journal = {Crystal Growth \& Design},
  volume = {9},
  number = {5},
  pages = {2525--2528},
  publisher = {American Chemical Society},
  issn = {1528-7483},
  doi = {10.1021/cg900065v},
  urldate = {2025-08-28}
}

@article{alonso-gonzalezLowDensityInAs2008,
  title = {Low Density {{InAs}} Quantum Dots with Control in Energy Emission and Top Surface Location},
  author = {{Alonso-Gonz{\'a}lez}, P. and Fuster, D. and Gonz{\'a}lez, L. and {Mart{\'i}n-S{\'a}nchez}, J. and Gonz{\'a}lez, Y.},
  year = 2008,
  month = nov,
  journal = {Applied Physics Letters},
  volume = {93},
  number = {18},
  pages = {183106},
  issn = {0003-6951},
  doi = {10.1063/1.3021070},
  urldate = {2025-08-28}
}

@article{ansaraBinaryDatabaseIII1994,
  title = {A Binary Database for {{III}}--{{V}} Compound Semiconductor Systems},
  author = {Ansara, I. and Chatillon, C. and Lukas, H. L. and Nishizawa, T. and Ohtani, H. and Ishida, K. and Hillert, M. and Sundman, B. and Argent, B. B. and Watson, A. and Chart, T. G. and Anderson, T.},
  year = 1994,
  month = apr,
  journal = {Calphad},
  volume = {18},
  number = {2},
  pages = {177--222},
  issn = {0364-5916},
  doi = {10.1016/0364-5916(94)90027-2},
  urldate = {2025-08-26}
}

@incollection{aspectSecondQuantumRevolution2023,
  title = {The {{Second Quantum Revolution}}: {{From Basic Concepts}} to {{Quantum Technologies}}},
  shorttitle = {The {{Second Quantum Revolution}}},
  booktitle = {Photonic {{Quantum Technologies}}},
  author = {Aspect, Alain},
  year = 2023,
  pages = {7--30},
  publisher = {John Wiley \& Sons, Ltd},
  doi = {10.1002/9783527837427.ch2},
  urldate = {2025-09-08},
  chapter = {2},
  copyright = {\copyright{} 2023 Wiley-VCH GmbH},
  isbn = {978-3-527-83742-7},
  langid = {english}
}

@article{atkinsonIndependentWavelengthDensity2012,
  title = {Independent Wavelength and Density Control of Uniform {{GaAs}}/{{AlGaAs}} Quantum Dots Grown by Infilling Self-Assembled Nanoholes},
  author = {Atkinson, P. and Zallo, E. and Schmidt, O. G.},
  year = 2012,
  month = sep,
  journal = {Journal of Applied Physics},
  volume = {112},
  number = {5},
  pages = {054303},
  issn = {0021-8979},
  doi = {10.1063/1.4748183},
  urldate = {2025-01-14}
}

@article{babinFullWaferProperty2022,
  title = {Full Wafer Property Control of Local Droplet Etched {{GaAs}} Quantum Dots},
  author = {Babin, Hans-Georg and Bart, Nikolai and Schmidt, Marcel and Spitzer, Nikolai and Wieck, Andreas D. and Ludwig, Arne},
  year = 2022,
  month = aug,
  journal = {Journal of Crystal Growth},
  volume = {591},
  pages = {126713},
  issn = {0022-0248},
  doi = {10.1016/j.jcrysgro.2022.126713},
  urldate = {2025-04-30}
}

@article{balesDynamicsIrreversibleIsland1994,
  title = {Dynamics of Irreversible Island Growth during Submonolayer Epitaxy},
  author = {Bales, G. S. and Chrzan, D. C.},
  year = 1994,
  month = sep,
  journal = {Physical Review B},
  volume = {50},
  number = {9},
  pages = {6057--6067},
  publisher = {American Physical Society},
  doi = {10.1103/PhysRevB.50.6057},
  urldate = {2025-07-18}
}

@article{bayerFineStructureNeutral2002,
  title = {Fine Structure of Neutral and Charged Excitons in Self-Assembled {{In}}({{Ga}}){{As}}/({{Al}}){{GaAs}} Quantum Dots},
  author = {Bayer, M. and Ortner, G. and Stern, O. and Kuther, A. and Gorbunov, A. A. and Forchel, A. and Hawrylak, P. and Fafard, S. and Hinzer, K. and Reinecke, T. L. and Walck, S. N. and Reithmaier, J. P. and Klopf, F. and Sch{\"a}fer, F.},
  year = 2002,
  month = may,
  journal = {Physical Review B},
  volume = {65},
  number = {19},
  pages = {195315},
  publisher = {American Physical Society},
  doi = {10.1103/PhysRevB.65.195315},
  urldate = {2025-01-15}
}

@article{bensonRegulatedEntangledPhotons2000,
  title = {Regulated and {{Entangled Photons}} from a {{Single Quantum Dot}}},
  author = {Benson, Oliver and Santori, Charles and Pelton, Matthew and Yamamoto, Yoshihisa},
  year = 2000,
  month = mar,
  journal = {Physical Review Letters},
  volume = {84},
  number = {11},
  pages = {2513--2516},
  publisher = {American Physical Society},
  doi = {10.1103/PhysRevLett.84.2513},
  urldate = {2025-01-15}
}

@article{biettiGalliumSurfaceDiffusion2014,
  title = {Gallium Surface Diffusion on {{GaAs}} (001) Surfaces Measured by Crystallization Dynamics of {{Ga}} Droplets},
  author = {Bietti, Sergio and Somaschini, Claudio and Esposito, Luca and Fedorov, Alexey and Sanguinetti, Stefano},
  year = 2014,
  month = sep,
  journal = {Journal of Applied Physics},
  volume = {116},
  number = {11},
  pages = {114311},
  issn = {0021-8979},
  doi = {10.1063/1.4895986},
  urldate = {2025-08-28}
}

@article{blackmanScalingBehaviorSubmonolayer1996,
  title = {Scaling Behavior in Submonolayer Film Growth: {{A}} One-Dimensional Model},
  shorttitle = {Scaling Behavior in Submonolayer Film Growth},
  author = {Blackman, J. A. and Mulheran, P. A.},
  year = 1996,
  month = oct,
  journal = {Physical Review B},
  volume = {54},
  number = {16},
  pages = {11681--11692},
  publisher = {American Physical Society},
  doi = {10.1103/PhysRevB.54.11681},
  urldate = {2025-07-18}
}

@article{braunSituTechniqueMeasuring1994,
  title = {In\hphantom{,}Situ Technique for Measuring {{Ga}} Segregation and Interface Roughness at {{GaAs}}/{{AlGaAs}} Interfaces},
  author = {Braun, W. and Ploog, K. H.},
  year = 1994,
  month = feb,
  journal = {Journal of Applied Physics},
  volume = {75},
  number = {4},
  pages = {1993--2001},
  issn = {0021-8979},
  doi = {10.1063/1.356324},
  urldate = {2025-08-12}
}

@article{caoLocalDropletEtching2022,
  title = {Local Droplet Etching on {{InAlAs}}/{{InP}} Surfaces with {{InAl}} Droplets},
  author = {Cao, Xin and Zhang, Yiteng and Ma, Chenxi and Wang, Yinan and Brechtken, Benedikt and Haug, Rolf J. and Rugeramigabo, Eddy P. and Zopf, Michael and Ding, Fei},
  year = 2022,
  month = may,
  journal = {AIP Advances},
  volume = {12},
  number = {5},
  pages = {055302},
  issn = {2158-3226},
  doi = {10.1063/5.0088012},
  urldate = {2025-08-26}
}

@article{cesuraDropletFreeSelfassembling2024,
  title = {Droplet Free Self-Assembling of High Density Nanoholes on {{GaAs}}(100) via Thermal Drilling},
  author = {Cesura, Federico and Vichi, Stefano and Tuktamyshev, Artur and Bietti, Sergio and Fedorov, Alexey and Sanguinetti, Stefano and Iizuka, Kanji and Tsukamoto, Shiro},
  year = 2024,
  month = mar,
  journal = {Journal of Crystal Growth},
  volume = {630},
  pages = {127588},
  issn = {0022-0248},
  doi = {10.1016/j.jcrysgro.2024.127588},
  urldate = {2025-09-10}
}

@article{chelluHighlyUniformGaSb2021,
  title = {Highly Uniform {{GaSb}} Quantum Dots with Indirect--Direct Bandgap Crossover at Telecom Range},
  author = {Chellu, Abhiroop and Hilska, Joonas and Penttinen, Jussi-Pekka and Hakkarainen, Teemu},
  year = 2021,
  month = may,
  journal = {APL Materials},
  volume = {9},
  number = {5},
  pages = {051116},
  issn = {2166-532X},
  doi = {10.1063/5.0049788},
  urldate = {2025-08-28}
}

@article{dasilvaGaAsQuantumDots2021,
  title = {{{GaAs}} Quantum Dots Grown by Droplet Etching Epitaxy as Quantum Light Sources},
  author = {{da Silva}, Saimon Filipe Covre and Undeutsch, Gabriel and Lehner, Barbara and Manna, Santanu and Krieger, Tobias M. and Reindl, Marcus and Schimpf, Christian and Trotta, Rinaldo and Rastelli, Armando},
  year = 2021,
  month = sep,
  journal = {Applied Physics Letters},
  volume = {119},
  number = {12},
  pages = {120502},
  issn = {0003-6951},
  doi = {10.1063/5.0057070},
  urldate = {2025-01-08}
}

@misc{dasilvaLowdensityInGaAsAlGaAs2025,
  title = {Low-Density {{InGaAs}}/{{AlGaAs Quantum Dots}} in {{Droplet-Etched Nanoholes}}},
  author = {{da Silva}, Saimon F. Covre and Garcia, Ailton J. and Aigner, Maximilian and Weidinger, Christian and Krieger, Tobias M. and Undeutsch, Gabriel and Deneke, Christoph and Bashir, Ishrat and Manna, Santanu and Peter, Melina and Brytavskyi, Ievgen and Aberl, Johannes and Rastelli, Armando},
  year = 2025,
  month = aug,
  number = {arXiv:2508.08400},
  eprint = {2508.08400},
  primaryclass = {physics},
  publisher = {arXiv},
  doi = {10.48550/arXiv.2508.08400},
  urldate = {2026-01-05},
  archiveprefix = {arXiv}
}

@article{deutschHarnessingPowerSecond2020,
  title = {Harnessing the {{Power}} of the {{Second Quantum Revolution}}},
  author = {Deutsch, Ivan H.},
  year = 2020,
  month = nov,
  journal = {PRX Quantum},
  volume = {1},
  number = {2},
  pages = {020101},
  publisher = {American Physical Society},
  doi = {10.1103/PRXQuantum.1.020101},
  urldate = {2026-01-15}
}

@article{deutschLocalDropletEtching2025,
  title = {Local Droplet Etching with {{In}}, {{Al}} and {{InAl}} in {{In0}}.{{52Al0}}.{{48As}} Layers for Generation of Quantum Dots Emitting in the Optical {{C-band}}},
  author = {Deutsch, D. and Zolatanosha, V. and Reuter, D.},
  year = 2025,
  month = oct,
  journal = {Journal of Crystal Growth},
  volume = {668},
  pages = {128247},
  issn = {0022-0248},
  doi = {10.1016/j.jcrysgro.2025.128247},
  urldate = {2026-01-05}
}

@article{deutschTelecomCbandPhoton2023,
  title = {Telecom {{C-band}} Photon Emission from ({{In}},{{Ga}}){{As}} Quantum Dots Generated by Filling Nanoholes in {{In0}}.{{52Al0}}.{{48As}} Layers},
  author = {Deutsch, D. and Buchholz, C. and Zolatanosha, V. and J{\"o}ns, K. D. and Reuter, D.},
  year = 2023,
  month = may,
  journal = {AIP Advances},
  volume = {13},
  number = {5},
  pages = {055009},
  issn = {2158-3226},
  doi = {10.1063/5.0147281},
  urldate = {2025-01-14}
}

@article{dowlingQuantumTechnologySecond2003,
  title = {Quantum Technology: The Second Quantum Revolution},
  shorttitle = {Quantum Technology},
  author = {Dowling, Jonathan P. and Milburn, Gerard J.},
  editor = {MacFarlane, A. G. J.},
  year = 2003,
  month = jun,
  journal = {Philosophical Transactions of the Royal Society A: Mathematical, Physical and Engineering Sciences},
  volume = {361},
  number = {1809},
  pages = {1655--1674},
  issn = {1364-503X},
  doi = {10.1098/rsta.2003.1227},
  urldate = {2026-01-15}
}

@article{fusterFundamentalRoleArsenic2014,
  title = {Fundamental Role of Arsenic Flux in Nanohole Formation by {{Ga}} Droplet Etching on {{GaAs}}(001)},
  author = {Fuster, David and Gonz{\'a}lez, Yolanda and Gonz{\'a}lez, Luisa},
  year = 2014,
  month = jun,
  journal = {Nanoscale Research Letters},
  volume = {9},
  number = {1},
  pages = {309},
  issn = {1556-276X},
  doi = {10.1186/1556-276X-9-309},
  urldate = {2025-01-15},
  langid = {english}
}

@article{gatosCharacteristics111Surfaces1960,
  title = {Characteristics of the \textbraceleft 111\textbraceright{} {{Surfaces}} of the {{III}}--{{V Intermetallic Compounds}}},
  author = {Gatos, Harry C. and Lavine, Mary C.},
  year = 1960,
  month = may,
  journal = {Journal of The Electrochemical Society},
  volume = {107},
  number = {5},
  pages = {427},
  publisher = {IOP Publishing},
  issn = {1945-7111},
  doi = {10.1149/1.2427712},
  urldate = {2025-08-19},
  langid = {english}
}

@article{gisinQuantumCryptography2002,
  title = {Quantum Cryptography},
  author = {Gisin, Nicolas and Ribordy, Gr{\'e}goire and Tittel, Wolfgang and Zbinden, Hugo},
  year = 2002,
  month = mar,
  journal = {Reviews of Modern Physics},
  volume = {74},
  number = {1},
  pages = {145--195},
  publisher = {American Physical Society},
  doi = {10.1103/RevModPhys.74.145},
  urldate = {2025-09-08}
}

@article{grafExcitonicStatesGaAs2014,
  title = {Excitonic States in {{GaAs}} Quantum Dots Fabricated by Local Droplet Etching},
  author = {Graf, A. and Sonnenberg, D. and Paulava, V. and Schliwa, A. and Heyn, {\relax Ch}. and Hansen, W.},
  year = 2014,
  month = mar,
  journal = {Physical Review B},
  volume = {89},
  number = {11},
  pages = {115314},
  publisher = {American Physical Society},
  doi = {10.1103/PhysRevB.89.115314},
  urldate = {2025-11-03}
}

@article{gurioliDropletEpitaxySemiconductor2019,
  title = {Droplet Epitaxy of Semiconductor Nanostructures for Quantum Photonic Devices},
  author = {Gurioli, Massimo and Wang, Zhiming and Rastelli, Armando and Kuroda, Takashi and Sanguinetti, Stefano},
  year = 2019,
  month = aug,
  journal = {Nature Materials},
  volume = {18},
  number = {8},
  pages = {799--810},
  publisher = {Nature Publishing Group},
  issn = {1476-4660},
  doi = {10.1038/s41563-019-0355-y},
  urldate = {2025-01-23},
  copyright = {2019 Springer Nature Limited},
  langid = {english}
}

@article{hallSolubilityIIICompound1963,
  title = {Solubility of {{III}}--{{V Compound Semiconductors}} in {{Column III Liquids}}},
  author = {Hall, R. N.},
  year = 1963,
  month = may,
  journal = {Journal of The Electrochemical Society},
  volume = {110},
  number = {5},
  pages = {385},
  publisher = {IOP Publishing},
  issn = {1945-7111},
  doi = {10.1149/1.2425770},
  urldate = {2025-11-13},
  langid = {english}
}

@article{heynDotSizeDependentExcitons2022,
  title = {Dot-{{Size Dependent Excitons}} in {{Droplet-Etched Cone-Shell GaAs Quantum Dots}}},
  author = {Heyn, Christian and Gr{\"a}fenstein, Andreas and Pirard, Geoffrey and Ranasinghe, Leonardo and Deneke, Kristian and Alshaikh, Ahmed and Bester, Gabriel and Hansen, Wolfgang},
  year = 2022,
  month = jan,
  journal = {Nanomaterials},
  volume = {12},
  number = {17},
  pages = {2981},
  publisher = {Multidisciplinary Digital Publishing Institute},
  issn = {2079-4991},
  doi = {10.3390/nano12172981},
  urldate = {2025-01-15},
  copyright = {http://creativecommons.org/licenses/by/3.0/},
  langid = {english}
}

@article{heynDynamicsMassTransport2015,
  title = {Dynamics of Mass Transport during Nanohole Drilling by Local Droplet Etching},
  author = {Heyn, Christian and Bartsch, Thorben and Sanguinetti, Stefano and Jesson, David and Hansen, Wolfgang},
  year = 2015,
  month = feb,
  journal = {Nanoscale Research Letters},
  volume = {10},
  number = {1},
  pages = {67},
  issn = {1556-276X},
  doi = {10.1186/s11671-015-0779-5},
  urldate = {2025-01-15},
  langid = {english}
}

@article{heynDynamicsSelfassembledDroplet2009,
  title = {Dynamics of Self-Assembled Droplet Etching},
  author = {Heyn, {\relax Ch}. and Stemmann, A. and Hansen, W.},
  year = 2009,
  month = oct,
  journal = {Applied Physics Letters},
  volume = {95},
  number = {17},
  pages = {173110},
  issn = {0003-6951},
  doi = {10.1063/1.3254216},
  urldate = {2025-07-24}
}

@article{heynFieldControlledQuantumDot2019,
  title = {Field-{{Controlled Quantum Dot}} to {{Ring Transformation}} in {{Wave-Function Tunable Cone-Shell Quantum Structures}}},
  author = {Heyn, Christian and K{\"u}ster, Achim and Zocher, Michael and Hansen, Wolfgang},
  year = 2019,
  journal = {physica status solidi (RRL) -- Rapid Research Letters},
  volume = {13},
  number = {1},
  pages = {1800245},
  issn = {1862-6270},
  doi = {10.1002/pssr.201800245},
  urldate = {2025-11-01},
  copyright = {\copyright{} 2018 WILEY-VCH Verlag GmbH \& Co. KGaA, Weinheim},
  langid = {english}
}

@article{heynHighlyUniformStrainfree2009,
  title = {Highly Uniform and Strain-Free {{GaAs}} Quantum Dots Fabricated by Filling of Self-Assembled Nanoholes},
  author = {Heyn, {\relax Ch}. and Stemmann, A. and K{\"o}ppen, T. and Strelow, {\relax Ch}. and Kipp, T. and Grave, M. and Mendach, S. and Hansen, W.},
  year = 2009,
  month = may,
  journal = {Applied Physics Letters},
  volume = {94},
  number = {18},
  pages = {183113},
  issn = {0003-6951},
  doi = {10.1063/1.3133338},
  urldate = {2025-01-13}
}

@article{heynInfluenceGaCoverage2009,
  title = {Influence of {{Ga}} Coverage and {{As}} Pressure on Local Droplet Etching of Nanoholes and Quantum Rings},
  author = {Heyn, {\relax Ch}. and Stemmann, A. and Eiselt, R. and Hansen, W.},
  year = 2009,
  month = mar,
  journal = {Journal of Applied Physics},
  volume = {105},
  number = {5},
  pages = {054316},
  issn = {0021-8979},
  doi = {10.1063/1.3079789},
  urldate = {2025-08-05}
}

@article{heynKineticModelLocal2011,
  title = {Kinetic Model of Local Droplet Etching},
  author = {Heyn, Christian},
  year = 2011,
  month = apr,
  journal = {Physical Review B},
  volume = {83},
  number = {16},
  pages = {165302},
  publisher = {American Physical Society},
  doi = {10.1103/PhysRevB.83.165302},
  urldate = {2025-01-13}
}

@article{heynModelingGaDroplet2021,
  title = {Modeling of {{Al}} and {{Ga Droplet Nucleation}} during {{Droplet Epitaxy}} or {{Droplet Etching}}},
  author = {Heyn, Christian and Feddersen, Stefan},
  year = 2021,
  month = feb,
  journal = {Nanomaterials},
  volume = {11},
  number = {2},
  pages = {468},
  publisher = {Multidisciplinary Digital Publishing Institute},
  issn = {2079-4991},
  doi = {10.3390/nano11020468},
  urldate = {2025-07-18},
  copyright = {http://creativecommons.org/licenses/by/3.0/},
  langid = {english}
}

@article{heynNanoholeFormationAlGaAs2009,
  title = {Nanohole Formation on {{AlGaAs}} Surfaces by Local Droplet Etching with Gallium},
  author = {Heyn, {\relax Ch}. and Stemmann, A. and Hansen, W.},
  year = 2009,
  month = mar,
  journal = {Journal of Crystal Growth},
  series = {International {{Conference}} on {{Molecular Beam Epitaxy}} ({{MBE-XV}})},
  volume = {311},
  number = {7},
  pages = {1839--1842},
  issn = {0022-0248},
  doi = {10.1016/j.jcrysgro.2008.11.001},
  urldate = {2025-07-24}
}

@article{heynOpticalPropertiesGaAs2009,
  title = {Optical {{Properties}} of {{GaAs Quantum Dots Fabricated}} by {{Filling}} of {{Self-Assembled Nanoholes}}},
  author = {Heyn, Ch and Stemmann, A. and K{\"o}ppen, T. and Strelow, Ch and Kipp, T. and Grave, M. and Mendach, S. and Hansen, W.},
  year = 2009,
  month = dec,
  journal = {Nanoscale Research Letters},
  volume = {5},
  number = {3},
  pages = {576},
  issn = {1556-276X},
  doi = {10.1007/s11671-009-9507-3},
  urldate = {2025-07-24},
  langid = {english}
}

@article{heynRegimesGaAsQuantum2007,
  title = {Regimes of {{GaAs}} Quantum Dot Self-Assembly by Droplet Epitaxy},
  author = {Heyn, {\relax Ch}. and Stemmann, A. and Schramm, A. and Welsch, H. and Hansen, W. and Nemcsics, {\'A}.},
  year = 2007,
  month = aug,
  journal = {Physical Review B},
  volume = {76},
  number = {7},
  pages = {075317},
  publisher = {American Physical Society},
  doi = {10.1103/PhysRevB.76.075317},
  urldate = {2025-08-28}
}

@article{heynRoleArsenicAluminum2016,
  title = {Role of {{Arsenic During Aluminum Droplet Etching}} of {{Nanoholes}} in {{AlGaAs}}},
  author = {Heyn, Christian and Zocher, Michel and Schn{\"u}ll, Sandra and Hansen, Wolfgang},
  year = 2016,
  month = sep,
  journal = {Nanoscale Research Letters},
  volume = {11},
  number = {1},
  pages = {428},
  issn = {1556-276X},
  doi = {10.1186/s11671-016-1648-6},
  urldate = {2025-01-15},
  langid = {english}
}

@article{heynScalingStructuralCharacteristics2014,
  title = {Scaling of the Structural Characteristics of Nanoholes Created by Local Droplet Etching},
  author = {Heyn, {\relax Ch}. and Schn{\"u}ll, S. and Hansen, W.},
  year = 2014,
  month = jan,
  journal = {Journal of Applied Physics},
  volume = {115},
  number = {2},
  pages = {024309},
  issn = {0021-8979},
  doi = {10.1063/1.4861722},
  urldate = {2025-07-29}
}

@article{heynSingledotSpectroscopyGaAs2010,
  title = {Single-Dot {{Spectroscopy}} of {{GaAs Quantum Dots Fabricated}} by {{Filling}} of {{Self-assembled Nanoholes}}},
  author = {Heyn, Ch and Klingbeil, M. and Strelow, Ch and Stemmann, A. and Mendach, S. and Hansen, W.},
  year = 2010,
  month = jul,
  journal = {Nanoscale Research Letters},
  volume = {5},
  number = {10},
  pages = {1633},
  issn = {1556-276X},
  doi = {10.1007/s11671-010-9687-x},
  urldate = {2025-07-24},
  langid = {english}
}

@article{hilskaNanoholeEtchingAlGaSb2021,
  title = {Nanohole {{Etching}} in {{AlGaSb}} with {{Gallium Droplets}}},
  author = {Hilska, Joonas and Chellu, Abhiroop and Hakkarainen, Teemu},
  year = 2021,
  month = apr,
  journal = {Crystal Growth \& Design},
  volume = {21},
  number = {4},
  pages = {1917--1923},
  publisher = {American Chemical Society},
  issn = {1528-7483},
  doi = {10.1021/acs.cgd.1c00113},
  urldate = {2025-08-28}
}

@article{huberHighlyIndistinguishableStrongly2017,
  title = {Highly Indistinguishable and Strongly Entangled Photons from Symmetric {{GaAs}} Quantum Dots},
  author = {Huber, Daniel and Reindl, Marcus and Huo, Yongheng and Huang, Huiying and Wildmann, Johannes S. and Schmidt, Oliver G. and Rastelli, Armando and Trotta, Rinaldo},
  year = 2017,
  month = may,
  journal = {Nature Communications},
  volume = {8},
  number = {1},
  pages = {15506},
  publisher = {Nature Publishing Group},
  issn = {2041-1723},
  doi = {10.1038/ncomms15506},
  urldate = {2025-11-01},
  copyright = {2017 The Author(s)},
  langid = {english}
}

@article{huberSemiconductorQuantumDots2018,
  title = {Semiconductor Quantum Dots as an Ideal Source of Polarization-Entangled Photon Pairs on-Demand: A Review},
  shorttitle = {Semiconductor Quantum Dots as an Ideal Source of Polarization-Entangled Photon Pairs on-Demand},
  author = {Huber, Daniel and Reindl, Marcus and Aberl, Johannes and Rastelli, Armando and Trotta, Rinaldo},
  year = 2018,
  month = jun,
  journal = {Journal of Optics},
  volume = {20},
  number = {7},
  pages = {073002},
  publisher = {IOP Publishing},
  issn = {2040-8986},
  doi = {10.1088/2040-8986/aac4c4},
  urldate = {2025-01-13},
  langid = {english}
}

@article{huberSingleparticlepictureBreakdownLaterally2019,
  title = {Single-Particle-Picture Breakdown in Laterally Weakly Confining {{GaAs}} Quantum Dots},
  author = {Huber, Daniel and Lehner, Barbara Ursula and Csontosov{\'a}, Diana and Reindl, Marcus and Schuler, Simon and {Covre da Silva}, Saimon Filipe and Klenovsk{\'y}, Petr and Rastelli, Armando},
  year = 2019,
  month = dec,
  journal = {Physical Review B},
  volume = {100},
  number = {23},
  pages = {235425},
  publisher = {American Physical Society},
  doi = {10.1103/PhysRevB.100.235425},
  urldate = {2025-11-03}
}

@article{huoUltrasmallExcitonicFine2013,
  title = {Ultra-Small Excitonic Fine Structure Splitting in Highly Symmetric Quantum Dots on {{GaAs}} (001) Substrate},
  author = {Huo, Y. H. and Rastelli, A. and Schmidt, O. G.},
  year = 2013,
  month = apr,
  journal = {Applied Physics Letters},
  volume = {102},
  number = {15},
  pages = {152105},
  issn = {0003-6951},
  doi = {10.1063/1.4802088},
  urldate = {2025-01-14}
}

@article{huoVolumeDependenceExcitonic2014,
  title = {Volume Dependence of Excitonic Fine Structure Splitting in Geometrically Similar Quantum Dots},
  author = {Huo, Y. H. and K{\v r}{\'a}pek, V. and Rastelli, A. and Schmidt, O. G.},
  year = 2014,
  month = jul,
  journal = {Physical Review B},
  volume = {90},
  number = {4},
  pages = {041304},
  publisher = {American Physical Society},
  doi = {10.1103/PhysRevB.90.041304},
  urldate = {2025-08-27}
}

@article{jaccodineUseModifiedFree1962,
  title = {Use of {{Modified Free Energy Theorems}} to {{Predict Equilibrium Growing}} and {{Etching Shapes}}},
  author = {Jaccodine, R. J.},
  year = 1962,
  month = aug,
  journal = {Journal of Applied Physics},
  volume = {33},
  number = {8},
  pages = {2643--2647},
  issn = {0021-8979},
  doi = {10.1063/1.1729036},
  urldate = {2025-08-19}
}

@article{jensenEffectMonomerEvaporation1997,
  title = {Effect of Monomer Evaporation on a Simple Model of Submonolayer Growth},
  author = {Jensen, Pablo and Larralde, Hern{\'a}n and Pimpinelli, Alberto},
  year = 1997,
  month = jan,
  journal = {Physical Review B},
  volume = {55},
  number = {4},
  pages = {2556--2569},
  publisher = {American Physical Society},
  doi = {10.1103/PhysRevB.55.2556},
  urldate = {2025-08-21}
}

@article{jonesEnergiesControllingNucleation1990,
  title = {Energies Controlling Nucleation and Growth Processes: {{The}} Case of {{Ag}}/{{W}}(110)},
  shorttitle = {Energies Controlling Nucleation and Growth Processes},
  author = {Jones, G. W. and Marcano, J. M. and N{\o}rskov, J. K. and Venables, J. A.},
  year = 1990,
  month = dec,
  journal = {Physical Review Letters},
  volume = {65},
  number = {26},
  pages = {3317--3320},
  publisher = {American Physical Society},
  doi = {10.1103/PhysRevLett.65.3317},
  urldate = {2025-09-08}
}

@article{jusserandLongRangeGallium1992,
  title = {Long Range Gallium Segregation in the {{AlAs}} Layers of {{GaAs}}/{{AlAs}} Superlattices},
  author = {Jusserand, Bernard and Mollot, Francis},
  year = 1992,
  month = jul,
  journal = {Applied Physics Letters},
  volume = {61},
  number = {4},
  pages = {423--425},
  issn = {0003-6951},
  doi = {10.1063/1.107902},
  urldate = {2025-08-12}
}

@article{kanjanachuchaiSelfRunningGaDroplets2013,
  title = {Self-{{Running Ga Droplets}} on {{GaAs}} (111){{A}} and (111){{B Surfaces}}},
  author = {Kanjanachuchai, Songphol and Euaruksakul, Chanan},
  year = 2013,
  month = aug,
  journal = {ACS Applied Materials \& Interfaces},
  volume = {5},
  number = {16},
  pages = {7709--7713},
  publisher = {American Chemical Society},
  issn = {1944-8244},
  doi = {10.1021/am402455u},
  urldate = {2025-09-06}
}

@article{keanGalliumDesorptionAlGaAs1991,
  title = {Gallium Desorption from ({{Al}},{{Ga}}){{As}} Grown by Molecular Beam Epitaxy at High Temperatures},
  author = {Kean, A. H. and Stanley, C. R. and Holland, M. C. and Martin, J. L. and Chapman, J. N.},
  year = 1991,
  month = may,
  journal = {Journal of Crystal Growth},
  volume = {111},
  number = {1},
  pages = {189--193},
  issn = {0022-0248},
  doi = {10.1016/0022-0248(91)90969-C},
  urldate = {2025-08-22}
}

@article{keilSolidstateEnsembleHighly2017,
  title = {Solid-State Ensemble of Highly Entangled Photon Sources at Rubidium Atomic Transitions},
  author = {Keil, Robert and Zopf, Michael and Chen, Yan and H{\"o}fer, Bianca and Zhang, Jiaxiang and Ding, Fei and Schmidt, Oliver G.},
  year = 2017,
  month = may,
  journal = {Nature Communications},
  volume = {8},
  number = {1},
  pages = {15501},
  publisher = {Nature Publishing Group},
  issn = {2041-1723},
  doi = {10.1038/ncomms15501},
  urldate = {2026-01-11},
  copyright = {2017 The Author(s)},
  langid = {english}
}

@article{kerbstDensityLimitsHigh2014,
  title = {Density Limits of High Temperature and Multiple Local Droplet Etching on {{AlAs}}},
  author = {Kerbst, J. and Heyn, {\relax Ch}. and Slobodskyy, T. and Hansen, W.},
  year = 2014,
  month = mar,
  journal = {Journal of Crystal Growth},
  volume = {389},
  pages = {18--22},
  issn = {0022-0248},
  doi = {10.1016/j.jcrysgro.2013.11.044},
  urldate = {2025-08-11}
}

@article{kleyNovelDiffusionMechanism1997,
  title = {Novel {{Diffusion Mechanism}} on the {{GaAs}}(001) {{Surface}}: {{The Role}} of {{Adatom-Dimer Interaction}}},
  shorttitle = {Novel {{Diffusion Mechanism}} on the {{GaAs}}(001) {{Surface}}},
  author = {Kley, Alexander and Ruggerone, Paolo and Scheffler, Matthias},
  year = 1997,
  month = dec,
  journal = {Physical Review Letters},
  volume = {79},
  number = {26},
  pages = {5278--5281},
  publisher = {American Physical Society},
  doi = {10.1103/PhysRevLett.79.5278},
  urldate = {2025-11-13}
}

@article{kruckCriticalAluminumEtch2024,
  title = {Critical {{Aluminum Etch Material Amount}} for {{Local Droplet-Etched Nanohole-Based GaAs Quantum Dots}}},
  author = {Kruck, Timo and Babin, Hans Georg and Wieck, Andreas D. and Ludwig, Arne},
  year = 2024,
  month = aug,
  journal = {Crystals},
  volume = {14},
  number = {8},
  pages = {714},
  publisher = {Multidisciplinary Digital Publishing Institute},
  issn = {2073-4352},
  doi = {10.3390/cryst14080714},
  urldate = {2025-04-29},
  copyright = {http://creativecommons.org/licenses/by/3.0/},
  langid = {english}
}

@article{kumahAtomicscaleMappingQuantum2009,
  title = {Atomic-Scale Mapping of Quantum Dots Formed by Droplet Epitaxy},
  author = {Kumah, Divine P. and Shusterman, Sergey and Paltiel, Yossi and Yacoby, Yizhak and Clarke, Roy},
  year = 2009,
  month = dec,
  journal = {Nature Nanotechnology},
  volume = {4},
  number = {12},
  pages = {835--838},
  publisher = {Nature Publishing Group},
  issn = {1748-3395},
  doi = {10.1038/nnano.2009.271},
  urldate = {2025-07-24},
  copyright = {2009 Springer Nature Limited},
  langid = {english}
}

@article{kusterDropletEtchingDeep2016,
  title = {Droplet Etching of Deep Nanoholes for Filling with Self-Aligned Complex Quantum Structures},
  author = {K{\"u}ster, Achim and Heyn, Christian and Ungeheuer, Arne and Juska, Gediminas and Tommaso Moroni, Stefano and Pelucchi, Emanuele and Hansen, Wolfgang},
  year = 2016,
  month = jun,
  journal = {Nanoscale Research Letters},
  volume = {11},
  number = {1},
  pages = {282},
  issn = {1556-276X},
  doi = {10.1186/s11671-016-1495-5},
  urldate = {2025-08-26},
  langid = {english}
}

@article{leeEvolutionSelfassembledSingle2006,
  title = {Evolution between Self-Assembled Single and Double Ring-like Nanostructures},
  author = {Lee, J H and Wang, Zh M and Abuwaar, Z Y and Strom, N W and Salamo, G J},
  year = 2006,
  month = jul,
  journal = {Nanotechnology},
  volume = {17},
  number = {15},
  pages = {3973},
  issn = {0957-4484},
  doi = {10.1088/0957-4484/17/15/061},
  urldate = {2025-08-11},
  langid = {english}
}

@article{leeSuperLowDensity2008,
  title = {Super {{Low Density InGaAs Semiconductor Ring-Shaped Nanostructures}}},
  author = {Lee, Jihoon H. and Wang, Zhiming M. and Ware, Morgan E. and Wijesundara, Kushal C. and Garrido, Mauricio and Stinaff, {\relax Eric}. A. and Salamo, Gregory J.},
  year = 2008,
  month = jun,
  journal = {Crystal Growth \& Design},
  volume = {8},
  number = {6},
  pages = {1945--1951},
  publisher = {American Chemical Society},
  issn = {1528-7483},
  doi = {10.1021/cg701263c},
  urldate = {2025-08-28}
}

@article{leguayUnveilingElectronicStructure2024,
  title = {Unveiling the Electronic Structure of {{GaSb}}/{{AlGaSb}} Quantum Dots Emitting in the Third Telecom Window},
  author = {Leguay, Lucie and Chellu, Abhiroop and Hilska, Joonas and Luna, Esperanza and Schliwa, Andrei and Guina, Mircea and Hakkarainen, Teemu},
  year = 2024,
  month = jan,
  journal = {Materials for Quantum Technology},
  volume = {4},
  number = {1},
  pages = {015401},
  publisher = {IOP Publishing},
  issn = {2633-4356},
  doi = {10.1088/2633-4356/ad207e},
  urldate = {2026-02-16},
  langid = {english}
}

@article{liangEnergyTransferUltralow2008,
  title = {Energy {{Transfer}} within {{Ultralow Density Twin InAs Quantum Dots Grown}} by {{Droplet Epitaxy}}},
  author = {Liang, Bao-Lai and Wang, Zhi-Ming and Wang, Xiao-Yong and Lee, Ji-Hoon and Mazur, Yuriy I. and Shih, Chih-Kang and Salamo, Gregory J.},
  year = 2008,
  month = nov,
  journal = {ACS Nano},
  volume = {2},
  number = {11},
  pages = {2219--2224},
  publisher = {American Chemical Society},
  issn = {1936-0851},
  doi = {10.1021/nn800224p},
  urldate = {2025-11-16}
}

@article{liHoledNanostructuresFormed2010,
  title = {Holed Nanostructures Formed by Aluminum Droplets on a {{GaAs}} Substrate},
  author = {Li, Alvason Zhenhua and Wang, Zhiming M. and Wu, Jiang and Salamo, Gregory J.},
  year = 2010,
  month = jul,
  journal = {Nano Research},
  volume = {3},
  number = {7},
  pages = {490--495},
  issn = {1998-0000},
  doi = {10.1007/s12274-010-0009-5},
  urldate = {2025-08-11},
  langid = {english}
}

@article{liOriginNanoholeFormation2014,
  title = {Origin of Nanohole Formation by Etching Based on Droplet Epitaxy},
  author = {Li, Xinlei and Wu, Jiang and Wang, Zhiming M. and Liang, Baolai and Lee, Jihoon and Kim, Eun-Soo and Salamo, Gregory J.},
  year = 2014,
  month = feb,
  journal = {Nanoscale},
  volume = {6},
  number = {5},
  pages = {2675--2681},
  publisher = {The Royal Society of Chemistry},
  issn = {2040-3372},
  doi = {10.1039/C3NR06064K},
  urldate = {2025-01-15},
  langid = {english}
}

@article{liSelectiveFormationMechanisms2013,
  title = {Selective Formation Mechanisms of Quantum Dots on Patterned Substrates},
  author = {Li, Xinlei},
  year = 2013,
  month = mar,
  journal = {Physical Chemistry Chemical Physics},
  volume = {15},
  number = {14},
  pages = {5238--5242},
  publisher = {The Royal Society of Chemistry},
  issn = {1463-9084},
  doi = {10.1039/C3CP43890B},
  urldate = {2026-01-15},
  langid = {english}
}

@article{liThermodynamicReassessmentAlAsGa2001,
  title = {A Thermodynamic Reassessment of the {{Al-As-Ga}} System},
  author = {Li, {\relax Ch}. and Li, J. -B. and Du, Z. and Lu, L. and Zhang, W.},
  year = 2001,
  month = jan,
  journal = {Journal of Phase Equilibria},
  volume = {22},
  number = {1},
  pages = {26--33},
  issn = {1054-9714},
  doi = {10.1361/105497101770339265},
  urldate = {2025-11-16},
  langid = {english}
}

@article{loblCorrelationsOpticalProperties2019,
  title = {Correlations between Optical Properties and {{Voronoi-cell}} Area of Quantum Dots},
  author = {L{\"o}bl, Matthias C. and Zhai, Liang and Jahn, Jan-Philipp and Ritzmann, Julian and Huo, Yongheng and Wieck, Andreas D. and Schmidt, Oliver G. and Ludwig, Arne and Rastelli, Armando and Warburton, Richard J.},
  year = 2019,
  month = oct,
  journal = {Physical Review B},
  volume = {100},
  number = {15},
  pages = {155402},
  publisher = {American Physical Society},
  doi = {10.1103/PhysRevB.100.155402},
  urldate = {2025-04-30}
}

@article{lyamkinaInvestigationIntermediateStage2010,
  title = {The {{Investigation}} of {{Intermediate Stage}} of {{Template Etching}} with {{Metal Droplets}} by {{Wetting Angle Analysis}} on (001) {{GaAs Surface}}},
  author = {Lyamkina, {\relax AA} and Dmitriev, {\relax DV} and Galitsyn, Yu G. and Kesler, {\relax VG} and Moshchenko, {\relax SP} and Toropov, {\relax AI}},
  year = 2010,
  month = sep,
  journal = {Nanoscale Res Lett},
  volume = {6},
  number = {1},
  pages = {42},
  issn = {1556-276X},
  doi = {10.1007/s11671-010-9790-z},
  urldate = {2025-08-28},
  langid = {english}
}

@article{massonEngineeringNanoholeEtchedQuantum2026,
  title = {Engineering {{Nanohole-Etched Quantum Dots}} for {{Telecom-Band Single-Photon Generation}}},
  author = {Masson, Ian M. and Hageman, Aden and Whittier, Caleb and Montealegre, David and Kamaliya, Bhaveshkumar and Bassim, Nabil D. and Prineas, John P. and Uppu, Ravitej},
  year = 2026,
  month = jan,
  journal = {ACS Nano},
  volume = {20},
  number = {3},
  pages = {2872--2880},
  publisher = {American Chemical Society},
  issn = {1936-0851},
  doi = {10.1021/acsnano.5c17982},
  urldate = {2026-02-04}
}

@article{michlStrainFreeGaSbQuantum2023,
  title = {Strain-{{Free GaSb Quantum Dots}} as {{Single-Photon Sources}} in the {{Telecom S-Band}}},
  author = {Michl, Johannes and Peniakov, Giora and Pfenning, Andreas and Hilska, Joonas and Chellu, Abhiroop and Bader, Andreas and Guina, Mircea and H{\"o}fling, Sven and Hakkarainen, Teemu and {Huber-Loyola}, Tobias},
  year = 2023,
  journal = {Advanced Quantum Technologies},
  volume = {6},
  number = {12},
  pages = {2300180},
  issn = {2511-9044},
  doi = {10.1002/qute.202300180},
  urldate = {2025-11-09},
  copyright = {\copyright{} 2023 The Authors. Advanced Quantum Technologies published by Wiley-VCH GmbH},
  langid = {english}
}

@article{mirandaBimodalIslandsizeDistributions2001,
  title = {Bimodal Island-Size Distributions in Submonolayer Growth},
  author = {Miranda, Rodolfo and Gallego, Jos{\'e} M.},
  year = 2001,
  month = aug,
  journal = {Physical Review B},
  volume = {64},
  number = {8},
  pages = {085426},
  publisher = {American Physical Society},
  doi = {10.1103/PhysRevB.64.085426},
  urldate = {2025-08-20}
}

@article{mulheranCaptureZonesScaling1996,
  title = {Capture Zones and Scaling in Homogeneous Thin-Film Growth},
  author = {Mulheran, P. A. and Blackman, J. A.},
  year = 1996,
  month = apr,
  journal = {Physical Review B},
  volume = {53},
  number = {15},
  pages = {10261--10267},
  publisher = {American Physical Society},
  doi = {10.1103/PhysRevB.53.10261},
  urldate = {2025-07-18}
}

@article{mulheranOriginsIslandSize1995,
  title = {The Origins of Island Size Scaling in Heterogeneous Film Growth},
  author = {Mulheran, P. A. and Blackman, J. A.},
  year = 1995,
  month = jul,
  journal = {Philosophical Magazine Letters},
  volume = {72},
  number = {1},
  pages = {55--60},
  publisher = {Taylor \& Francis},
  issn = {0950-0839},
  doi = {10.1080/09500839508241614},
  urldate = {2025-07-18}
}

@article{nemcsicsCrosssectionalTransmissionElectron2011,
  title = {Cross-Sectional Transmission Electron Microscopy of {{GaAs}} Quantum Dots Fabricated by Filling of Droplet-Etched Nanoholes},
  author = {Nemcsics, {\'A}. and Heyn, {\relax Ch}. and T{\'o}th, L. and Dobos, L. and Stemmann, A. and Hansen, W.},
  year = 2011,
  month = nov,
  journal = {Journal of Crystal Growth},
  volume = {335},
  number = {1},
  pages = {58--61},
  issn = {0022-0248},
  doi = {10.1016/j.jcrysgro.2011.09.005},
  urldate = {2025-11-03}
}

@article{nguyenEnhancedElectronSpinCoherence2023,
  title = {Enhanced {{Electron-Spin Coherence}} in a {{GaAs Quantum Emitter}}},
  author = {Nguyen, Giang N. and Spinnler, Clemens and Hogg, Mark R. and Zhai, Liang and Javadi, Alisa and Schrader, Carolin A. and Erbe, Marcel and Wyss, Marcus and Ritzmann, Julian and Babin, Hans-Georg and Wieck, Andreas D. and Ludwig, Arne and Warburton, Richard J.},
  year = 2023,
  month = nov,
  journal = {Physical Review Letters},
  volume = {131},
  number = {21},
  pages = {210805},
  publisher = {American Physical Society},
  doi = {10.1103/PhysRevLett.131.210805},
  urldate = {2026-02-16}
}

@article{nothernTemplatedependentNucleationMetallic2012,
  title = {Template-Dependent Nucleation of Metallic Droplets},
  author = {Nothern, Denis M. and Millunchick, Joanna M.},
  year = 2012,
  month = sep,
  journal = {Journal of Vacuum Science \& Technology B},
  volume = {30},
  number = {6},
  pages = {060603},
  issn = {2166-2746},
  doi = {10.1116/1.4754563},
  urldate = {2025-07-18}
}

@article{ohtakeExtremelyHighLowDensity2015,
  title = {Extremely {{High-}} and {{Low-Density}} of {{Ga Droplets}} on {{GaAs}}\textbraceleft 111\textbraceright{{A}},{{B}}: {{Surface-Polarity Dependence}}},
  shorttitle = {Extremely {{High-}} and {{Low-Density}} of {{Ga Droplets}} on {{GaAs}}\textbraceleft 111\textbraceright{{A}},{{B}}},
  author = {Ohtake, Akihiro and Ha, Neul and Mano, Takaaki},
  year = 2015,
  month = jan,
  journal = {Crystal Growth \& Design},
  volume = {15},
  number = {1},
  pages = {485--488},
  publisher = {American Chemical Society},
  issn = {1528-7483},
  doi = {10.1021/cg501545n},
  urldate = {2025-08-22}
}

@article{orieuxSemiconductorDevicesEntangled2017,
  title = {Semiconductor Devices for Entangled Photon Pair Generation: A Review},
  shorttitle = {Semiconductor Devices for Entangled Photon Pair Generation},
  author = {Orieux, Adeline and Versteegh, Marijn A M and J{\"o}ns, Klaus D and Ducci, Sara},
  year = 2017,
  month = may,
  journal = {Reports on Progress in Physics},
  volume = {80},
  number = {7},
  pages = {076001},
  publisher = {IOP Publishing},
  issn = {0034-4885},
  doi = {10.1088/1361-6633/aa6955},
  urldate = {2025-09-08},
  langid = {english}
}

@article{pimpinelliCaptureZoneScalingIsland2007,
  title = {Capture-{{Zone Scaling}} in {{Island Nucleation}}: {{Universal Fluctuation Behavior}}},
  shorttitle = {Capture-{{Zone Scaling}} in {{Island Nucleation}}},
  author = {Pimpinelli, Alberto and Einstein, T. L.},
  year = 2007,
  month = nov,
  journal = {Physical Review Letters},
  volume = {99},
  number = {22},
  pages = {226102},
  publisher = {American Physical Society},
  doi = {10.1103/PhysRevLett.99.226102},
  urldate = {2025-07-18}
}

@article{pimpinelliPimpinelliEinsteinReply2010,
  title = {Pimpinelli and {{Einstein Reply}}:},
  shorttitle = {Pimpinelli and {{Einstein Reply}}},
  author = {Pimpinelli, Alberto and Einstein, T. L.},
  year = 2010,
  month = apr,
  journal = {Physical Review Letters},
  volume = {104},
  number = {14},
  pages = {149602},
  publisher = {American Physical Society},
  doi = {10.1103/PhysRevLett.104.149602},
  urldate = {2025-07-21}
}

@incollection{pimpinelliScalingCrossoversModels1999,
  title = {Scaling and Crossovers in Models for Thin Film Growth},
  booktitle = {Morphological {{Organization}} in {{Epitaxial Growth}} and {{Removal}}},
  author = {Pimpinelli, Alberto and Jensen, Pablo and Larralde, Hern{\'a}n and Peyla, Philippe},
  year = 1999,
  month = jan,
  series = {Series on {{Directions}} in {{Condensed Matter Physics}}},
  volume = {Volume 14},
  pages = {121--148},
  publisher = {WORLD SCIENTIFIC},
  doi = {10.1142/9789812816245_0006},
  urldate = {2025-01-14},
  isbn = {978-981-02-3471-3}
}

@article{pimpinelliScalingExponentEqualities2014,
  title = {Scaling and {{Exponent Equalities}} in {{Island Nucleation}}: {{Novel Results}} and {{Application}} to {{Organic Films}}},
  shorttitle = {Scaling and {{Exponent Equalities}} in {{Island Nucleation}}},
  author = {Pimpinelli, Alberto and Tumbek, Levent and Winkler, Adolf},
  year = 2014,
  month = mar,
  journal = {The Journal of Physical Chemistry Letters},
  volume = {5},
  number = {6},
  pages = {995--998},
  publisher = {American Chemical Society},
  doi = {10.1021/jz500282t},
  urldate = {2025-07-21}
}

@article{raabOswaldRipeningShape2000,
  title = {Oswald Ripening and Shape Transitions of Self-Assembled {{PbSe}} Quantum Dots on {{PbTe}} (111) during Annealing},
  author = {Raab, A. and Springholz, G.},
  year = 2000,
  month = nov,
  journal = {Applied Physics Letters},
  volume = {77},
  number = {19},
  pages = {2991--2993},
  issn = {0003-6951},
  doi = {10.1063/1.1323733},
  urldate = {2025-05-05}
}

@article{ratschNucleationTheoryEarly2003,
  title = {Nucleation Theory and the Early Stages of Thin Film Growth},
  author = {Ratsch, C. and Venables, J. A.},
  year = 2003,
  month = sep,
  journal = {Journal of Vacuum Science \& Technology A},
  volume = {21},
  number = {5},
  pages = {S96-S109},
  issn = {0734-2101},
  doi = {10.1116/1.1600454},
  urldate = {2025-09-02}
}

@article{reyesUnifiedModelDroplet2013,
  title = {Unified Model of Droplet Epitaxy for Compound Semiconductor Nanostructures: {{Experiments}} and Theory},
  shorttitle = {Unified Model of Droplet Epitaxy for Compound Semiconductor Nanostructures},
  author = {Reyes, Kristofer and Smereka, Peter and Nothern, Denis and Millunchick, Joanna Mirecki and Bietti, Sergio and Somaschini, Claudio and Sanguinetti, Stefano and Frigeri, Cesare},
  year = 2013,
  month = apr,
  journal = {Physical Review B},
  volume = {87},
  number = {16},
  pages = {165406},
  publisher = {American Physical Society},
  doi = {10.1103/PhysRevB.87.165406},
  urldate = {2025-11-11}
}

@article{rubensteinSolubilitiesGaAsMetallic1966,
  title = {Solubilities of {{GaAs}} in {{Metallic Solvents}}},
  author = {Rubenstein, M.},
  year = 1966,
  month = jul,
  journal = {Journal of The Electrochemical Society},
  volume = {113},
  number = {7},
  pages = {752},
  publisher = {IOP Publishing},
  issn = {1945-7111},
  doi = {10.1149/1.2424107},
  urldate = {2026-01-20},
  langid = {english}
}

@article{sablonStructuralEvolutionFormation2008,
  title = {Structural {{Evolution During Formation}} and {{Filling}} of {{Self-patterned Nanoholes}} on {{GaAs}} (100) {{Surfaces}}},
  author = {Sablon, {\relax KA} and Wang, Zh M. and Salamo, {\relax GJ} and Zhou, Lin and Smith, David J.},
  year = 2008,
  month = nov,
  journal = {Nanoscale Research Letters},
  volume = {3},
  number = {12},
  pages = {530},
  issn = {1556-276X},
  doi = {10.1007/s11671-008-9194-5},
  urldate = {2025-08-12},
  langid = {english}
}

@article{salaInAsInPQuantum2020,
  title = {{{InAs}}/{{InP Quantum Dots}} in {{Etched Pits}} by {{Droplet Epitaxy}} in {{Metalorganic Vapor Phase Epitaxy}}},
  author = {Sala, Elisa Maddalena and Na, Young In and Godsland, Max and Trapalis, Aristotelis and Heffernan, Jon},
  year = 2020,
  journal = {physica status solidi (RRL) -- Rapid Research Letters},
  volume = {14},
  number = {8},
  pages = {2000173},
  issn = {1862-6270},
  doi = {10.1002/pssr.202000173},
  urldate = {2025-08-26},
  copyright = {\copyright{} 2020 WILEY-VCH Verlag GmbH \& Co. KGaA, Weinheim},
  langid = {english}
}

@article{salaLocalDropletEtching2024,
  title = {Local {{Droplet Etching}} with {{Indium Droplets}} on {{InP}}(100) by {{Metal}}--{{Organic Vapor Phase Epitaxy}}},
  author = {Sala, Elisa Maddalena and In Na, Young and Heffernan, Jon},
  year = 2024,
  month = nov,
  journal = {Crystal Growth \& Design},
  volume = {24},
  number = {22},
  pages = {9571--9580},
  publisher = {American Chemical Society},
  issn = {1528-7483},
  doi = {10.1021/acs.cgd.4c01097},
  urldate = {2025-08-26}
}

@incollection{sanguinettiDropletEpitaxyNanostructures2018,
  title = {Droplet {{Epitaxy}} of {{Nanostructures}}},
  booktitle = {Molecular {{Beam Epitaxy}}},
  author = {Sanguinetti, Stefano and Bietti, Sergio and Koguchi, Nobuyuki},
  year = 2018,
  month = jan,
  pages = {293--314},
  publisher = {Elsevier},
  doi = {10.1016/B978-0-12-812136-8.00013-X},
  urldate = {2025-11-16},
  langid = {american}
}

@article{santoriTriggeredSinglePhotons2001,
  title = {Triggered {{Single Photons}} from a {{Quantum Dot}}},
  author = {Santori, Charles and Pelton, Matthew and Solomon, Glenn and Dale, Yseulte and Yamamoto, Yoshihisa},
  year = 2001,
  month = feb,
  journal = {Physical Review Letters},
  volume = {86},
  number = {8},
  pages = {1502--1505},
  publisher = {American Physical Society},
  doi = {10.1103/PhysRevLett.86.1502},
  urldate = {2025-11-17}
}

@article{senellartHighperformanceSemiconductorQuantumdot2017,
  title = {High-Performance Semiconductor Quantum-Dot Single-Photon Sources},
  author = {Senellart, Pascale and Solomon, Glenn and White, Andrew},
  year = 2017,
  month = nov,
  journal = {Nature Nanotechnology},
  volume = {12},
  number = {11},
  pages = {1026--1039},
  publisher = {Nature Publishing Group},
  issn = {1748-3395},
  doi = {10.1038/nnano.2017.218},
  urldate = {2025-01-08},
  copyright = {2017 Springer Nature Limited},
  langid = {english}
}

@article{shorlinShapeCycleGa2007,
  title = {Shape Cycle of {{Ga}} Clusters on {{GaAs}} during Coalescence Growth},
  author = {Shorlin, K. and {Zinke-Allmang}, M.},
  year = 2007,
  month = jun,
  journal = {Surface Science},
  volume = {601},
  number = {12},
  pages = {2438--2444},
  issn = {0039-6028},
  doi = {10.1016/j.susc.2007.04.019},
  urldate = {2025-08-19}
}

@article{solomonEffectsMonolayerCoverage1995,
  title = {Effects of Monolayer Coverage, Flux Ratio, and Growth Rate on the Island Density of {{InAs}} Islands on {{GaAs}}},
  author = {Solomon, G. S. and Trezza, J. A. and Harris, Jr., J. S.},
  year = 1995,
  month = jun,
  journal = {Applied Physics Letters},
  volume = {66},
  number = {23},
  pages = {3161--3163},
  issn = {0003-6951},
  doi = {10.1063/1.113709},
  urldate = {2025-01-13}
}

@article{solomonSubstrateTemperatureMonolayer1995,
  title = {Substrate Temperature and Monolayer Coverage Effects on Epitaxial Ordering of {{InAs}} and {{InGaAs}} Islands on {{GaAs}}},
  author = {Solomon, G. S. and Trezza, J. A. and Harris, Jr., J. S.},
  year = 1995,
  month = feb,
  journal = {Applied Physics Letters},
  volume = {66},
  number = {8},
  pages = {991--993},
  issn = {0003-6951},
  doi = {10.1063/1.113822},
  urldate = {2025-01-13}
}

@article{solomonVerticallyAlignedElectronically1996,
  title = {Vertically {{Aligned}} and {{Electronically Coupled Growth Induced InAs Islands}} in {{GaAs}}},
  author = {Solomon, G. S. and Trezza, J. A. and Marshall, A. F. and Harris, J. S., {\relax Jr}.},
  year = 1996,
  month = feb,
  journal = {Physical Review Letters},
  volume = {76},
  number = {6},
  pages = {952--955},
  publisher = {American Physical Society},
  doi = {10.1103/PhysRevLett.76.952},
  urldate = {2026-01-09}
}

@article{sonnenbergHighlyVersatileUltralow2012,
  title = {Highly Versatile Ultra-Low Density {{GaAs}} Quantum Dots Fabricated by Filling of Self-Assembled Nanoholes},
  author = {Sonnenberg, D. and Graf, A. and Paulava, V. and Hansen, W. and Heyn, {\relax Ch}.},
  year = 2012,
  month = oct,
  journal = {Applied Physics Letters},
  volume = {101},
  number = {14},
  pages = {143106},
  issn = {0003-6951},
  doi = {10.1063/1.4756945},
  urldate = {2025-01-14}
}

@article{spitzerTelecomOBandQuantum2024,
  title = {Telecom {{O-Band Quantum Dots Fabricated}} by {{Droplet Etching}}},
  author = {Spitzer, Nikolai and Kersting, Elias and Grell, Meret and Kohminaei, Danial and Schmidt, Marcel and Bart, Nikolai and Wieck, Andreas D. and Ludwig, Arne},
  year = 2024,
  month = dec,
  journal = {Crystals},
  volume = {14},
  number = {12},
  pages = {1014},
  publisher = {Multidisciplinary Digital Publishing Institute},
  issn = {2073-4352},
  doi = {10.3390/cryst14121014},
  urldate = {2025-01-14},
  copyright = {http://creativecommons.org/licenses/by/3.0/},
  langid = {english}
}

@article{springthorpeMeasurementGaAsSurface1987,
  title = {Measurement of {{GaAs}} Surface Oxide Desorption Temperatures},
  author = {SpringThorpe, A. J. and Ingrey, S. J. and Emmerstorfer, B. and Mandeville, P. and Moore, W. T.},
  year = 1987,
  month = jan,
  journal = {Applied Physics Letters},
  volume = {50},
  number = {2},
  pages = {77--79},
  issn = {0003-6951},
  doi = {10.1063/1.97824},
  urldate = {2025-05-13}
}

@article{stemmannLocalDropletEtching2008,
  title = {Local Droplet Etching of Nanoholes and Rings on {{GaAs}} and {{AlGaAs}} Surfaces},
  author = {Stemmann, A. and Heyn, {\relax Ch}. and K{\"o}ppen, T. and Kipp, T. and Hansen, W.},
  year = 2008,
  month = sep,
  journal = {Applied Physics Letters},
  volume = {93},
  number = {12},
  pages = {123108},
  issn = {0003-6951},
  doi = {10.1063/1.2981517},
  urldate = {2025-07-24}
}

@article{stemmannLocalEtchingNanoholes2009,
  title = {Local Etching of Nanoholes and Quantum Rings with {{InxGa1}}-x Droplets},
  author = {Stemmann, A. and K{\"o}ppen, T. and Grave, M. and Wildfang, S. and Mendach, S. and Hansen, W. and Heyn, {\relax Ch}.},
  year = 2009,
  month = sep,
  journal = {Journal of Applied Physics},
  volume = {106},
  number = {6},
  pages = {064315},
  issn = {0021-8979},
  doi = {10.1063/1.3225759},
  urldate = {2025-07-24}
}

@article{stevensSurfaceDiffusionMeasurements2017,
  title = {Surface Diffusion Measurements of {{In}} on {{InGaAs}} Enabled by Droplet Epitaxy},
  author = {Stevens, Margaret A. and Tomasulo, Stephanie and Maximenko, Sergey and Vandervelde, Thomas E. and Yakes, Michael K.},
  year = 2017,
  month = may,
  journal = {Journal of Applied Physics},
  volume = {121},
  number = {19},
  pages = {195302},
  issn = {0021-8979},
  doi = {10.1063/1.4983257},
  urldate = {2025-08-28}
}

@article{stinaffOpticalSignaturesCoupled2006,
  title = {Optical {{Signatures}} of {{Coupled Quantum Dots}}},
  author = {Stinaff, E. A. and Scheibner, M. and Bracker, A. S. and Ponomarev, I. V. and Korenev, V. L. and Ware, M. E. and Doty, M. F. and Reinecke, T. L. and Gammon, D.},
  year = 2006,
  month = feb,
  journal = {Science},
  volume = {311},
  number = {5761},
  pages = {636--639},
  publisher = {American Association for the Advancement of Science},
  doi = {10.1126/science.1121189},
  urldate = {2026-01-09}
}

@article{stockillQuantumDotSpin2016,
  title = {Quantum Dot Spin Coherence Governed by a Strained Nuclear Environment},
  author = {Stockill, R. and Le Gall, C. and Matthiesen, C. and Huthmacher, L. and Clarke, E. and Hugues, M. and Atat{\"u}re, M.},
  year = 2016,
  month = sep,
  journal = {Nature Communications},
  volume = {7},
  number = {1},
  pages = {12745},
  publisher = {Nature Publishing Group},
  issn = {2041-1723},
  doi = {10.1038/ncomms12745},
  urldate = {2026-01-04},
  copyright = {2016 The Author(s)},
  langid = {english}
}

@article{tersoffRunningDropletsGallium2009,
  title = {Running {{Droplets}} of {{Gallium}} from {{Evaporation}} of {{Gallium Arsenide}}},
  author = {Tersoff, J. and Jesson, D. E. and Tang, W. X.},
  year = 2009,
  month = apr,
  journal = {Science},
  volume = {324},
  number = {5924},
  pages = {236--238},
  publisher = {American Association for the Advancement of Science},
  doi = {10.1126/science.1169546},
  urldate = {2025-09-06}
}

@article{tuktamyshevLocalDropletEtching2024,
  title = {Local Droplet Etching of a Vicinal {{InGaAs}}(111){{A}} Metamorphic Layer},
  author = {Tuktamyshev, Artur and Lambardi, Davide and Vichi, Stefano and Cesura, Federico and Cecchi, Stefano and Fedorov, Alexey and Bietti, Sergio and Sanguinetti, Stefano},
  year = 2024,
  month = oct,
  journal = {Applied Surface Science},
  volume = {669},
  pages = {160450},
  issn = {0169-4332},
  doi = {10.1016/j.apsusc.2024.160450},
  urldate = {2025-08-27}
}

@article{tuktamyshevNucleationGaDroplets2021,
  title = {Nucleation of {{Ga}} Droplets Self-Assembly on {{GaAs}}(111){{A}} Substrates},
  author = {Tuktamyshev, Artur and Fedorov, Alexey and Bietti, Sergio and Vichi, Stefano and Tambone, Riccardo and Tsukamoto, Shiro and Sanguinetti, Stefano},
  year = 2021,
  month = mar,
  journal = {Scientific Reports},
  volume = {11},
  number = {1},
  pages = {6833},
  publisher = {Nature Publishing Group},
  issn = {2045-2322},
  doi = {10.1038/s41598-021-86339-3},
  urldate = {2025-08-21},
  copyright = {2021 The Author(s)},
  langid = {english}
}

@article{tuktamyshevTemperatureActivatedDimensionality2019,
  title = {Temperature {{Activated Dimensionality Crossover}} in the {{Nucleation}} of {{Quantum Dots}} by {{Droplet Epitaxy}} on {{GaAs}}(111){{A Vicinal Substrates}}},
  author = {Tuktamyshev, Artur and Fedorov, Alexey and Bietti, Sergio and Tsukamoto, Shiro and Sanguinetti, Stefano},
  year = 2019,
  month = oct,
  journal = {Scientific Reports},
  volume = {9},
  number = {1},
  pages = {14520},
  publisher = {Nature Publishing Group},
  issn = {2045-2322},
  doi = {10.1038/s41598-019-51161-5},
  urldate = {2025-08-22},
  copyright = {2019 The Author(s)},
  langid = {english}
}

@article{vasilenkoFormationGaAsNanostructures2015,
  title = {Formation of {{GaAs}} Nanostructures by Droplet Epitaxy---{{Monte Carlo}} Simulation},
  author = {Vasilenko, Maxim A. and Neizvestny, Igor G. and Shwartz, Nataliya L.},
  year = 2015,
  month = may,
  journal = {Computational Materials Science},
  volume = {102},
  pages = {286--292},
  issn = {0927-0256},
  doi = {10.1016/j.commatsci.2015.02.032},
  urldate = {2025-11-11}
}

@article{venablesAtomicProcessesCrystal1994,
  title = {Atomic Processes in Crystal Growth},
  author = {Venables, John A},
  year = 1994,
  month = jan,
  journal = {Surface Science},
  volume = {299--300},
  pages = {798--817},
  issn = {0039-6028},
  doi = {10.1016/0039-6028(94)90698-X},
  urldate = {2025-01-24}
}

@book{venablesIntroductionSurfaceThin2000,
  title = {Introduction to {{Surface}} and {{Thin Film Processes}}},
  author = {Venables, John A.},
  year = 2000,
  publisher = {Cambridge University Press},
  address = {Cambridge},
  doi = {10.1017/CBO9780511755651},
  urldate = {2025-09-02},
  isbn = {978-0-521-78500-6}
}

@article{venablesNucleationGrowthThin1984,
  title = {Nucleation and Growth of Thin Films},
  author = {Venables, J. A. and Spiller, G. D. T. and Hanbucken, M.},
  year = 1984,
  month = apr,
  journal = {Reports on Progress in Physics},
  volume = {47},
  number = {4},
  pages = {399},
  issn = {0034-4885},
  doi = {10.1088/0034-4885/47/4/002},
  urldate = {2025-01-14},
  langid = {english}
}

@article{venablesRateEquationApproaches1973,
  title = {Rate Equation Approaches to Thin Film Nucleation Kinetics},
  author = {Venables, J. A.},
  year = 1973,
  month = mar,
  journal = {The Philosophical Magazine: A Journal of Theoretical Experimental and Applied Physics},
  volume = {27},
  number = {3},
  pages = {697--738},
  publisher = {Taylor \& Francis},
  issn = {0031-8086},
  doi = {10.1080/14786437308219242},
  urldate = {2025-08-20}
}

@article{vonkFacetingLocalDropletetched2018,
  title = {Faceting of Local Droplet-Etched Nanoholes in {{AlGaAs}}},
  author = {Vonk, Vedran and Slobodskyy, Taras and Keller, Thomas F. and Richard, Marie-Ingrid and Fern{\'a}ndez, Sara and Schulli, Tobias and Heyn, Christian and Hansen, Wolfgang and Stierle, Andreas},
  year = 2018,
  month = oct,
  journal = {Physical Review Materials},
  volume = {2},
  number = {10},
  pages = {106001},
  publisher = {American Physical Society},
  doi = {10.1103/PhysRevMaterials.2.106001},
  urldate = {2025-08-19}
}

@article{wacaserPreferentialInterfaceNucleation2009,
  title = {Preferential {{Interface Nucleation}}: {{An Expansion}} of the {{VLS Growth Mechanism}} for {{Nanowires}}},
  shorttitle = {Preferential {{Interface Nucleation}}},
  author = {Wacaser, Brent A. and Dick, Kimberly A. and Johansson, Jonas and Borgstr{\"o}m, Magnus T. and Deppert, Knut and Samuelson, Lars},
  year = 2009,
  journal = {Advanced Materials},
  volume = {21},
  number = {2},
  pages = {153--165},
  issn = {1521-4095},
  doi = {10.1002/adma.200800440},
  urldate = {2025-11-12},
  copyright = {Copyright \copyright{} 2009 WILEY-VCH Verlag GmbH \& Co. KGaA, Weinheim},
  langid = {english}
}

@article{waltonNucleationVaporDeposits1962,
  title = {Nucleation of {{Vapor Deposits}}},
  author = {Walton, D.},
  year = 1962,
  month = nov,
  journal = {The Journal of Chemical Physics},
  volume = {37},
  number = {10},
  pages = {2182--2188},
  issn = {0021-9606},
  doi = {10.1063/1.1732985},
  urldate = {2025-07-21}
}

@article{wangMechanismAluminumDroplet2023,
  title = {Mechanism of {{Aluminum Droplet Nucleation}} and {{Ripening}} on {{GaAs}}(001) {{Surface}} by {{Molecular Beam Epitaxy}}},
  author = {Wang, Yi and Jiang, Chong and Huang, Yanbin and Ding, Zhao and Luo, Zijiang and Wang, Jihong and Guo, Xiang},
  year = 2023,
  month = jan,
  journal = {Journal of Electronic Materials},
  volume = {52},
  number = {1},
  pages = {463--470},
  issn = {1543-186X},
  doi = {10.1007/s11664-022-10012-2},
  urldate = {2025-08-12},
  langid = {english}
}

@article{wangNanoholesFabricatedSelfassembled2007,
  title = {Nanoholes Fabricated by Self-Assembled Gallium Nanodrill on {{GaAs}}(100)},
  author = {Wang, {\relax Zh}. M. and Liang, B. L. and Sablon, K. A. and Salamo, G. J.},
  year = 2007,
  month = mar,
  journal = {Applied Physics Letters},
  volume = {90},
  number = {11},
  pages = {113120},
  issn = {0003-6951},
  doi = {10.1063/1.2713745},
  urldate = {2025-01-13}
}

@article{wuFabricationUltralowdensityQuantum2017,
  title = {Fabrication of Ultralow-Density Quantum Dots by Droplet Etching Epitaxy},
  author = {Wu, Jiang and Wang, Zhiming M. and Li, Xinlei and Mazur, Yuriy I. and Salamo, Gregory J.},
  year = 2017,
  month = nov,
  journal = {Journal of Materials Research},
  volume = {32},
  number = {21},
  pages = {4095--4101},
  issn = {2044-5326},
  doi = {10.1557/jmr.2017.408},
  urldate = {2025-08-27},
  langid = {english}
}

@article{xingTransientStableProfiles2017,
  title = {Transient and {{Stable Profiles During Anisotropic Wet Etching}} of {{Quartz}}},
  author = {Xing, Yan and Gos{\'a}lvez, Miguel A. and Zhang, Hui and Li, Yuan and Qiu, Xiaoli},
  year = 2017,
  month = oct,
  journal = {Journal of Microelectromechanical Systems},
  volume = {26},
  number = {5},
  pages = {1063--1072},
  issn = {1941-0158},
  doi = {10.1109/JMEMS.2017.2707096},
  urldate = {2025-08-19}
}

@article{youngImprovedFidelityTriggered2006,
  title = {Improved Fidelity of Triggered Entangled Photons from Single Quantum Dots},
  author = {Young, Robert J and Stevenson, R Mark and Atkinson, Paola and Cooper, Ken and Ritchie, David A and Shields, Andrew J},
  year = 2006,
  month = feb,
  journal = {New Journal of Physics},
  volume = {8},
  number = {2},
  pages = {29},
  issn = {1367-2630},
  doi = {10.1088/1367-2630/8/2/029},
  urldate = {2025-09-08},
  langid = {english}
}

@article{yuHighlyUniformSymmetric2019,
  title = {Highly Uniform and Symmetric Epitaxial {{InAs}} Quantum Dots Embedded inside {{Indium}} Droplet Etched Nanoholes},
  author = {Yu, Ying and Zhong, Hancheng and Yang, Jiawei and Liu, Lin and Liu, Jin and Yu, Siyuan},
  year = 2019,
  month = sep,
  journal = {Nanotechnology},
  volume = {30},
  number = {48},
  pages = {485001},
  publisher = {IOP Publishing},
  issn = {0957-4484},
  doi = {10.1088/1361-6528/ab3efb},
  urldate = {2025-08-27},
  langid = {english}
}

@article{zhaiQuantumInterferenceIdentical2022,
  title = {Quantum Interference of Identical Photons from Remote {{GaAs}} Quantum Dots},
  author = {Zhai, Liang and Nguyen, Giang N. and Spinnler, Clemens and Ritzmann, Julian and L{\"o}bl, Matthias C. and Wieck, Andreas D. and Ludwig, Arne and Javadi, Alisa and Warburton, Richard J.},
  year = 2022,
  month = aug,
  journal = {Nature Nanotechnology},
  volume = {17},
  number = {8},
  pages = {829--833},
  publisher = {Nature Publishing Group},
  issn = {1748-3395},
  doi = {10.1038/s41565-022-01131-2},
  urldate = {2025-01-17},
  copyright = {2022 The Author(s), under exclusive licence to Springer Nature Limited},
  langid = {english}
}

@article{zinke-allmangClusteringSurfaces1992,
  title = {Clustering on Surfaces},
  author = {{Zinke-Allmang}, Martin and Feldman, Leonard C. and Grabow, Marcia H.},
  year = 1992,
  month = dec,
  journal = {Surface Science Reports},
  volume = {16},
  number = {8},
  pages = {377--463},
  issn = {0167-5729},
  doi = {10.1016/0167-5729(92)90006-W},
  urldate = {2025-09-06}
}

@article{zocherAlloyingLocalDroplet2019,
  title = {Alloying during Local Droplet Etching of {{AlGaAs}} Surfaces with Aluminium},
  author = {Zocher, M. and Heyn, {\relax Ch}. and Hansen, W.},
  year = 2019,
  month = jan,
  journal = {Journal of Applied Physics},
  volume = {125},
  number = {2},
  pages = {025306},
  issn = {0021-8979},
  doi = {10.1063/1.5053464},
  urldate = {2025-01-15}
}
\end{document}